
%
%
%

\def\CC{{\bf C}}
\def\crnt#1{$J^{(K)}(#1)$}
\def\lK{$\lambda_K$}
\def\JK{$J^{(K)}$}
\def\AK{${\cal A}^{(K)}$}
\def\sc{structure constant}
\def\ca{chiral algebra}
\def\FS{fractional supersymmetry}
\def\hat{\widehat}
\def\tilde{\widetilde}
\def\brak#1#2{\left[\matrix{#2\cr #1\cr}\right]}

\input phyzzx
\overfullrule=0pt

\Pubnum={CLNS 91/1059}
\date{April 1991 (revised)}
\pubtype={ }
\titlepage
\bigskip
\title{\bf Structure Constants of the Fractional Supersymmetry
Chiral Algebras}
\author{Philip C. Argyres\foot{pca@strange.tn.cornell.edu,
pca@crnlnuc.bitnet}, James M. Grochocinski
\foot{james@beauty.tn.cornell.edu, james@crnlnuc.bitnet},
and S.-H. Henry Tye}
\address{Newman Laboratory of Nuclear Studies \break
Cornell University \break
Ithaca, N.Y. 14853-5001}
\abstract
{The \FS\ \ca s, \AK, in two-dimensional conformal
field theory are extended Virasoro algebras with fractional
spin currents \JK.  We show that associativity and closure of
\AK\ determines its \sc s in the case that the Virasoro
algebra is extended by precisely one current.
We compute the structure constants
of the \AK\ algebras explicitly and we show that correlators
of \JK's satisfy non-Abelian braiding relations.}
\endpage

\REF\ZamW{A.B.~Zamolodchikov \journal Theor. Math. Phys.&65
(85) 1205.}
\REF\Wn{V.A.~Fateev and S.L.~Lykyanov {\journal Int. J. Mod.
Phys.&A3 (88) 507;} A.~Bilal and J.-L.~Gervais {\journal Phys.
Lett.&206B (88) 412} \journal Nucl. Phys.&B318 (89) 579.}
\REF\ZFPara{A.B.~Zamolodchikov and V.A.~Fateev \journal Sov.
Phys. J.E.T.P.& 62 (85) 215.}
\REF\ZFft{V.A.~Fateev and A.B.~Zamolodchikov \journal Theor.
Math. Phys.& 71 (88) 451.}
\REF\Kastor{D.~Kastor, E.~Martinec and Z.~Qiu \journal Phys.
Lett.&200B (88) 434.}
\REF\Bagger{J.~Bagger, D.~Nemeschansky and S.~Yankielowicz
\journal Phys.  Rev. Lett.&60 (88) 389.}
\REF\Rava{F.~Ravanini \journal Mod. Phys. Lett.&A3 (88) 397.}
\REF\Felder {G.~Felder \journal Nucl. Phys.&B317 (89) 215.}
\REF\CLT {S.~Chung, E.~Lyman and S.-H.~H.~Tye, {\it ``Fractional
Supersymmetry and Minimal Coset Models in Conformal Field Theory,''}
Cornell preprint CLNS 91/1057.}
\REF\ACT {C.~Ahn, S.~Chung and S.-H.~H.~Tye, {\it ``New
Parafermion, SU(2) Coset, and N=2 Superconformal Field Theories,''}
Cornell preprint CLNS 91/1053.}
\REF\Zam {A.B.~Zamolodchikov, LOMI preprint, Lectures in Beijing
Summer School, 1989.}
\REF\ABL {C.~Ahn, D.~Bernard and A.~LeClair {\journal Nucl.
Phys.&B346 (90) 409,} and references therein.}
\REF\ALT {P.~Argyres, A.~LeClair and S.-H.~H.~Tye \journal
Phys. Lett.&253B (91) 306.}
\REF\ArT {P.~Argyres and S.-H.H.~Tye, Cornell preprint CLNS 91/1068,
{\it ``Fractional Superstrings with Space-Time Critical Dimensions
Four and Six,''} July 1991.}
\REF\ZFWZW {A.B.~Zamolodchikov and V.A.~Fateev \journal Sov. J.
Nucl. Phys.& 43 (86) 657.}
\REF\Dots {Vl.S.~Dotsenko {\journal Nucl. Phys.&B338 (90) 747;} Landau
Inst. preprint, October, 1990.}
\REF\ZPo {A.B.~Zamolodchikov and R.~Poghossian \journal Sov. J. Nucl.
Phys.&47 (88) 1461.}
\REF\Pog {R.~Poghossian \journal Int. J. Mod. Phys.&6A (91) 2005.}
\REF\MooSigh{G.~Moore and N.~Seiberg {\journal Phys. Lett.&212B
(88) 451}
{\journal Nucl. Phys.&B313 (89) 16} \journal Commun. Math.
Phys.&123 (89) 177.}
\REF\FFKI{G.~Felder, J.~Fr\"{o}hlich and G.~Keller {\journal
Commun. Math.  Phys.&124 (89) 417} \journal Commun. Math.
Phys.&130 (90) 1.}
\REF\BPZ {A.A. Belavin, A.M. Polyakov and A.B. Zamolodchikov,
Nucl. Phys.  {\bf B241} (1984), 333.}
\REF\Waki{M.~Wakimoto \journal Commun. Math. Phys.&104 (86) 604.}
\REF\DotFat {Vl.S.~Dotsenko and V.A.~Fateev {\journal Nucl.
Phys.&B240 (84) 312} \journal Nucl.Phys.&B251 (85) 691.}
\REF\FFK {G.~Felder, J.~Fr\"{o}hlich and G.~Keller \journal
Commun. Math.  Phys.&124 (89) 647.}
\REF\FeiF{B.L.~Feigin and D.B.~Fuchs {\journal Funct. Anal.
Appl.&16 (82) 114;} G.~Felder \journal Nucl. Phys.&B317 (89) 215.}
\REF\BerLe{D.~Bernard and A.~LeClair \journal Phys. Lett.&247B (90) 309.}
\REF\FQS{D.~Friedan, Z.~Qiu and S.~Shenker \journal Phys. Rev.
Lett.&52 (84) 1575.}
\REF\CIZ{A.~Cappelli, C.~Itzykson and J.-B.~Zuber \journal
Nucl. Phys.&B280 (87) 445.}
\REF\RavMod{F.~Ravanini \journal Mod. Phys. Lett.&A3 (88) 271.}
\REF\Gott {See {\it e.g.}~K.~Gottfried, {\it Quantum Mechanics
Volume I: Fundamentals,} Benjamin-Cummings, Reading,
Massachusetts (1966).}
\TABLE\tone{Realizations of known representations of the
\AK\ \FS\ \ca\ in terms of ``free'' fields.  $(\varphi,\alpha_0)$
stands for a boson with background charge, and $(\omega,\omega^+)$
the $c=2$ first order bosonic free field system.}
\FIG\fone{The action of fusion transformations $\alpha$ on
conformal blocks.}
\FIG\ftwo{The action of braiding transformations $\beta$ on
conformal blocks.}
\FIG\fthree{Associativity constraint for the $\phi_{3,1}$ four-point
correlation function in the $K=4$ minimal model.}
\FIG\ffour{The action of twisting transformations $\gamma$
on conformal blocks.}
\FIG\ffive{Illustration of the relation $\beta=\alpha\gamma\alpha$ (5.2)
relating the braiding transformation to two fusions and a twist.}
\FIG\fsix{Illustration of the relation $\beta=\gamma\alpha\gamma$ (A12)
relating the braiding transformation to two twists and a fusion.}

\chapter{Introduction}

Extended chiral algebras (algebras which include the Virasoro algebra
as a subalgebra) are the main tool for organizing representations and
classifying models in two-dimensional conformal field theory (CFT).
They also constitute the basis of attempts to form new ``string-like''
theories by gauging these chiral algebras.  For both CFT and string
applications, the \ca s must be associative operator algebras.  This
condition places tight constraints on the structure constants and
operator content of the extended Virasoro algebra.  The \sc s of
many \ca s have been determined in this way [\ZamW,\Wn,\ZFPara,\ZFft].
The goal of this paper is to do the same for the \FS\ \ca s.
These are non-local algebras which enlarge the Virasoro
symmetry algebra to include a fractional spin current.
In quantum field theory, it is well known that fractional spin
fields satisfy non-trivial braiding relations.
In the two-dimensional \FS\ \ca s that we will consider,
some of the currents have scaling dimensions other than (half)
integers.  Since the currents are chiral fields, the
fact that they have fractional dimensions implies that they also have
fractional spins.  The correlators of these currents obey
non-Abelian braiding relations.  In this paper we show how such
braiding relations can be disentangled.

There is a large body of evidence that \FS\ \ca s organize infinite
sets of CFTs in the same way that local extended Virasoro algebras do.
These algebras were conjectured [\Kastor,\Bagger,\Rava] in the context
of $SU(2)$ coset constructions, and were presented as a new way of
organizing $1<c<3$ representations in CFT.  Using a BRST cohomology
approach [\Felder] based on the \FS\ algebras, some details of
the coset models have been worked out [\CLT], while new coset
models have been constructed [\ACT].  These new coset models
are $SU(2)_K\otimes SU(2)_L/SU(2)_{K+L}$ where both $K$ and $L$
are rational; use of the \FS\ \ca s enables one to construct their
branching functions.  Another body of evidence comes from the
study of perturbed CFT and massive integrable systems, where
\FS\ \ca s play an important role [\Zam,\ABL].  More recently,
the \FS\ \ca s have also been proposed [\ALT,\ArT] as the basis
for fractional superstrings, generalizing the bosonic string and
the superstring.  For all these applications it is important
to determine the \sc s of the \FS\ \ca s---indeed, without them
the \ca s and their representation theory are not fully defined.

In general, the \FS\ \ca s are generated by the energy-momentum
tensor and a set of fractional spin currents $J_i(z)$.  For the
$K$-th \FS\ algebra, the $J_i$ must have dimension $(K+4)/(K+2)$.
In addition, closure of the operator algebra may require that
chiral fields with other fractional dimensions enter in the \FS\ \ca.
In this paper, we will assume that there is only one fractional
spin current.  This is not necessarily the most interesting case,
but serves as an example of our approach to the problem of
constructing \FS\ \ca s.  Our approach is equally applicable to
the more general case, although the analysis clearly becomes
more involved.

The fractional supersymmetry algebras will be denoted \AK. The
fractional spin of the currents is reflected in the appearance
of fractional powers in the \AK\ operator product expansions (OPEs).
They are
generated by two currents: the energy-momentum tensor $T(z)$ and the
fractional supersymmetry current \crnt{z}, with the following OPEs
  $$\eqalign{
  T(z)T(w)= & {c\over 2}(z-w)^{-4}\left\{1+\ldots\right\},\cr
  T(z)J^{(K)}(w)= & \Delta (z-w)^{-2}\left\{J^{(K)}(w)+\ldots\right\},\cr
  J^{(K)}(z)J^{(K)}(w)= & (z-w)^{-2\Delta}\left\{1+\ldots\right\}
  +\lambda_K(c)(z-w)^{-\Delta}\left\{J^{(K)}(w)+\ldots\right\},\cr}
  \eqn\Zi
  $$
where $\Delta = (K+4)/(K+2)$ is the dimension of \JK, and where
the dots represent the Virasoro descendants. The existence of the
current $J^{(K)}$ was proposed earlier [\Kastor,\Bagger,\Rava].
However, the existence of the algebra \Zi\ can be established
only if the coupling \lK\ can be consistently determined.  The
main result of this paper is to show that the fractional
supersymmetry algebra is consistent with \lK\ given by
  $$
  \lambda^2_K (c) = {{2 K^2 (c_{111})^2}\over{3(K+4)(K+2)}}
  \left[{{3 (K+4)^2}\over{K(K+2)}}{1\over c} - 1\right]
  \eqn\Zii
  $$
where, with $\rho = {1\over{K+2}}$,
  $$
  (c_{111})^2=
  {{\sin^2(\pi\rho)\sin^2(4\pi\rho)}\over
  {\sin^3(2\pi\rho)\sin(3\pi\rho)}}
  {{\Gamma^3(\rho)\Gamma^2(4\rho)}\over
  {\Gamma(3\rho)\Gamma^4(2\rho)}}
  \eqn\Zia
  $$
is the {\it chiral} structure constant for three spin-$1$ primaries
in the $SU(2)_K$ Wess-Zumino-Witten (WZW) theory.
The result \Zii\ was obtained earlier [\ALT]
but without determining $c_{111}$. It is an important point of this
paper that $c_{111}$ does not equal the usual $SU(2)_K$ structure
constants computed in non-chiral theories combining left- and
right-moving sectors [\ZFWZW,\Dots].  Notice that \lK\ vanishes
when $K=2$; this corresponds to the superVirasoro algebra where
the supercurrent \JK\ has dimension $3/2$.  Also, for $K=4$, we
recover the known structure constant for (a diagonal version of)
the spin-$4/3$ algebra [\ZFft].

At this point the reader may worry that if the above chiral algebras
have different structure constants than those required in a non-chiral
theory, then how can these chiral algebras organize non-chiral (local)
CFTs?  Indeed, the existence of
non-local chiral algebras in two-dimensional CFT
is problematical in general because
the locality of the full (left plus right)
correlation functions rules out the appearance of the fractional spin
currents as fields in the full CFT.  Nevertheless, it is found that the
Ward identities following from such chiral algebras, even though they
involve chiral structure constants
and unphysical correlation functions with
cuts on the world-sheet, do relate physical correlation functions of
non-chiral theories, and thus serve as organizing symmetries of local
CFTs.  The apparent contradiction
between the chiral and non-chiral structure
constants is avoided because the fractional spin current associated with
the chiral structure constants never
appears in physical correlators, but only
in the derivation of the Ward identities relating physical correlators.
The structure constants and correlation functions for sets of CFTs
have been worked out in this way for the two simplest \FS\ \ca s---the
superVirasoro algebra and the spin-$4/3$ algebra [\ZPo,\ZFft,\Pog].
The evidence from $SU(2)$ coset models and from two-dimensional
integrable models, mentioned above, suggests that a similar picture
should be true of the other \FS\ \ca s.
This paper is intended as a first step in the exploration of fractional
spin \ca s.  In particular, we will not be able to throw much light on
precisely how a local (non-chiral) two-dimensional CFT is interpreted
as a representation of a non-local \ca.  Instead, we will concentrate
on the construction and properties of the \ca s themselves, without
addressing the problem of their representation theory.

The \lK\ found in \Zii\ is determined
by imposing two consistency conditions,
namely associativity and closure.
By the closure condition, we mean that
only the identity and \JK\ and their
Virasoro descendants are allowed in the
chiral algebra:  no new primary
operators with different dimensions can
appear in \Zi, even with positive
fractional powers of $(z-w)$, since such
terms introduce new cuts in the $z$-plane.
The determination of \lK\ for the \AK\ algebras \Zi\ can be thought of
as a continuation of the program initiated by Zamolodchikov [\ZamW], in
which one picks the chiral operator content for a proposed chiral
algebra, and calculates which values of the structure constants and
central charge (if any) are allowed by associativity.
The non-local property of these new algebras makes the
analysis more involved.

The associativity constraints on the \sc s of any \ca\
are derived from demanding invariance of the correlation functions of
the chiral currents under different ways of grouping the operators.
This procedure has been formulated in terms of the braiding and fusion
matrices of conformal blocks [\MooSigh,\FFKI]. Associativity is
equivalent to invariance of correlation functions under fusion
transformations, while invariance under braiding transformations
corresponds to demanding monodromy
invariance (locality).  The associativity
constraints are usually enlarged to
include all constraints derived from
invariance of correlation functions
under the group of all transformations
generated by the braiding and fusion
matrices.  However, these formulations
assume half-integral dimensions for
the chiral algebra currents, and so
must be reexamined in the case of
fractional spin \ca s.  In particular, we
have found that imposing invariance
of the chiral correlators under
the full transformation group is too strong a condition:  in general
there are no solutions for the \sc s. Demanding invariance only under
associativity (fusion) transformations
in general fixes the \sc s completely.

To determine the associativity
constraints, one would like to calculate
the four-point function of \JK\
currents using a Ward identity derived
from the $J^{(K)}J^{(K)}$ OPE.
However, for general fractional dimension
of the current, $\Delta$, there
are two different cuts, $(z-w)^{-2\Delta}$
and $(z-w)^{-\Delta}$, appearing in the $J^{(K)}J^{(K)}$ OPE.  This
situation prevents us from performing the analytic continuation in $z$
necessary to derive a useful Ward identity.  An equivalent way of
describing this problem is in terms of the fusion matrix for \JK\
four-point conformal blocks.  The fusion matrix reflects the structure
of cuts on the complex plane, and satisfies complicated nonlinear
constraints following from conformal invariance ({\it e.g.} the pentagon
identity, \etc\ of ref.~[\MooSigh]).  Unfortunately, nontrivial
solutions to these constraints for fields with fractional dimensions
are known only in the specific cases of the $c<1$ minimal models,
and the $SU(2)_K$ WZW theories.  Therefore, to
make further progress in understanding the \FS\ algebras \AK, we
find it easiest to relate them to the minimal and WZW theories.

To this end, we will briefly review the known representations of the
\AK\ algebras and describe their free field realizations, as summarized
in Table~\tone.  The \AK\ algebras were first suggested
[\Kastor,\Bagger,\Rava] as current algebras in the
$SU(2)_K\otimes SU(2)_L/SU(2)_{K+L}$ coset models.  These models can,
for fixed $K$, be constructed by a generalized Feigin-Fuchs procedure
involving a $Z_K$-parafermion and a boson with background charge
depending on $L$.  The current \JK\ is determined by the requirement
that it commutes with the screening charges in the parafermion plus
boson construction.  Its dimension is found to be $\Delta=(K+4)/(K+2)$,
and its explicit form in terms of $Z_K$-parafermion fields has been
computed (see eqn.~(3.7) below) [\ALT].  The $L=1$ case corresponds
to the unitary minimal models with $c=c_{\rm min}=1-6/(K+2)(K+3)$,
and \JK\ reduces to the $\phi_{3,1}$ primary field in the BPZ [\BPZ]
classification of minimal model fields.  When $L$ is taken to infinity
the background charge goes to zero and one recovers the $SU(2)_K$ WZW
model.  In this limit, the fractional supersymmetry current can be
expressed as $J^{(K)}(z)=q_{ab}J^a_{-1}\Phi_{(1)}^b(z)$,
where $\Phi_{(1)}^b(z)$
is the adjoint representation WZW primary field of dimension $2/(K+2)$,
$J^a_{-1}$ are the dimension $1$ modes of the Kac-Moody currents, and
$q_{ab}$ is the Killing form for $SU(2)$.

The strategy for calculating \lK\ is as follows.  The fusion
matrices have been computed for the $SU(2)_K$ WZW models [\Dots] using
the Wakimoto free-field realization [\Waki].  Thus we can implement
the associativity and closure conditions for the $SU(2)_K$
representations of \AK\ in order to determine \lK\ at $c=c_{SU(2)}$.
Since \JK\ is a WZW descendant of the spin-$1$ primary field,
\lK\ is proportional to the $c_{111}$ chiral $SU(2)_K$ structure
constant.
The closure condition corresponds to the requirement
that $c_{112}$ (the \sc\ for two spin-$1$ fields to fuse to
a spin-$2$ field) is zero.
The generalized Feigin-Fuchs realization of the
coset representations of \AK\ gives an expression for \JK\ valid for
central charges $c_{\rm min}\leq c\leq c_{SU(2)}$.
By computing the $J^{(K)}J^{(K)}$ OPE using this expression, in which
the background charge dependence is explicit, we can determine the
central charge dependence of \lK.
Coupled with the $SU(2)_K$ WZW model
calculation of the value of \lK\ at
$c=c_{SU(2)}$, this determines \lK\ at all $c$.

Alternatively, we can determine \lK\ using the fusion matrices of the
minimal model representations of \AK.  These have been calculated in
[\DotFat,\FFK] using the Feigin-Fuchs realization of the minimal
series in terms of a free boson with background charge [\FeiF].
Here \JK\ is identified with the $\phi_{3,1}$ Virasoro
primary field.  The
closure condition corresponds to the decoupling of the $\phi_{5,1}$
primary field from the chiral algebra.  Then the associativity argument
determines \lK\ at the particular value of the central charge
$c_{\rm min}$, and hence provides an independent check on our
calculations.

Since most readers are more familiar with the minimal model formalism,
we will present the logic of chiral algebra associativity and the
calculation of $\lambda_K(c_{\rm min})$ first, in Section~2.
Section~3 then describes the determination of the central charge
dependence of \lK\ using the generalized Feigin-Fuchs realization
of \AK, and in Section~4 we outline the technically somewhat more
complicated associativity argument for the $SU(2)_K$ representations.
We collect in appendices some old and new
results for fusion and braiding matrices
in the minimal and WZW models,
and for $SU(2)_K$ Ward identities
useful in calculating parafermion OPEs.

In Section~5 we discuss various issues related to the fractional
supersymmetry algebras and to non-local chiral algebras in general.
Since the fractional supersymmetry algebras are non-local, they obey
non-trivial braiding properties.
We show that the $J^{(K)}$ conformal blocks
obey non-Abelian braiding properties
that are closely related to the fractional spin
of the current.  It is interesting to
ask if other non-local chiral algebras
can be constructed along the lines of the \AK\ algebras presented
above.  In fact, the spin-$4/3$
parafermion algebra of [\ZFft] is closely
related to the ${\cal A}^{(4)}$ algebra, except that it has two
spin-$4/3$ currents instead of one.  The relation between these two
algebras suggests new fractional
supersymmetry algebras associated with
$SU(2)$.  Finally, we comment on the
possibility of using fields with
dimensions other than those appearing in \Zi\ to generate new
non-local chiral algebras.


\chapter{Associativity Constraints on Chiral Algebras}

To compute the structure constants
of a chiral operator algebra, we follow
the bootstrap procedure of Belavin, Polyakov,
and Zamolodchikov [\BPZ].  In particular
we will express the four-point correlation
functions of the theory in question
in terms of the conformal blocks and
then translate the associativity of
the operator product algebra into
conditions involving the blocks and the
structure constants.  In principle the
blocks are known (in practice only
for the minimal series and $SU(2)_K$ models),
so the associativity condition
gives constraints on the structure constants.
We will see that for chiral
algebras these constraints are sufficient
to solve for the structure
constants.  Also, locality (monodromy
invariance) of the correlation function
is too strong a condition to impose:
in general there are no solutions
satisfying both associativity and locality.

For concreteness, we illustrate this idea in the context of the
minimal unitary models, though it should be clear that the basic idea
is more general.  We want to calculate in the minimal
unitary series with an enlarged chiral algebra. In particular, in addition
to the energy-momentum tensor, suppose that the $\phi_{3,1}$ primary
field is also a chiral algebra current.  We take the $\phi_{3,1}$ field to
be a left-moving (holomorphic) field and we consider the four-point
chiral correlation function
  $$
  G(z_i) = \langle \phi_{3,1}(z_1) \phi_{3,1}(z_2)
  \phi_{3,1}(z_3) \phi_{3,1}(z_4) \rangle.
  \eqn\Ii
  $$
{}From the minimal model fusion rules
  $$
  \phi_{m,1}\times\phi_{n,1} = \sum_{k=|m-n|+1}^{{\rm min}\{m+n-1,~
  2K+3-m-n\} }\phi_{k,1}
  \eqn\IIxxva
  $$
where $k$ is stepped by two in the sum,
the $\phi_{3,1}$ fields satisfy the OPE
  $$\eqalign{
  \phi_{3,1}(z) \phi_{3,1}(w) = & (z-w)^{-2\Delta}d_{331}[1] +
  (z-w)^{-\Delta}d_{333}[\phi_{3,1}(w)] \cr
  & + (z-w)^{\Delta+1}d_{335}[\phi_{5,1}(w)].\cr}
  \eqn\Iii
  $$
The square brackets denote the primary field and its descendants as well
as their $(z-w)$ dependence.  $\Delta$ is the conformal weight of the
$\phi_{3,1}$ field.  The $d_{331}$ structure constant can be set to one,
fixing the normalization of the $\phi_{3,1}$ field; we will do so
henceforth.  Note that the correlation \Ii\ is symbolic of the more
precise expression involving Feigin-Fuchs conjugate fields and
screening charges, as described in [\DotFat].
Parameterizing the minimal model central charge by
  $$
  c = 1-{6\over{(K+2)(K+3)}}
  \eqn\IIi
  $$
for $K=1,2,...$, we have $\Delta=(K+4)/(K+2)$ for $K\geq 2.$
Strictly speaking, the $\phi_{5,1}$ field only exists for $K\geq 4$;
we will consider the cases $K=2$ and $K=3$ at the end of this Section.
Since the $\phi_{5,1}$
fields appear in the $\phi_{3,1}$ OPE, it too must be a chiral field.
Note, however, that since its dimension is $\Delta_{5,1}=3\Delta+1$,
it appears with a positive fractional power of $(z-w)$ in the
$\phi_{3,1}$ OPE.

We impose associativity by requiring that $G(z_i)$
be the same independent of the order in which the $\phi_{3,1}$ fields are
fused. For example, as $z_1 \rightarrow z_2$ we can replace
$\phi_{3,1}(z_1) \phi_{3,1}(z_2)$ in $G(z_i)$ by its OPE.  Then we
only have three-point functions left in the determination of $G(z_i)$, whose
form is fixed by SL(2,\CC) invariance and whose normalizations are the
structure constants of the chiral algebra. Choosing the normalization
$d_{331}=1$, we find
  $$
  G(z_i) = F_3 (z_i) +
  {(d_{333})}^2 F_2 (z_i) +
  {(d_{335})}^2 F_1 (z_i)
  \eqn\Iiii
  $$
where the $F_k (z_i)$ are the holomorphic conformal blocks.  These
blocks must be normalized so that, as $z_1\rightarrow z_2$ the
expansion of the $F_k$'s in powers of $z_{12}=z_1-z_2$ starts
with unity, up to the $z_i$-dependence required by $SL(2,\CC)$
invariance.  For the $\phi_{3,1}$ conformal blocks, for example,
this means
  $$\eqalign{
  F_1(z_i) = & (z_{12}z_{34})^{\Delta+1}(z_{13}z_{24})^{-3\Delta-1}
  \left\{1+{\cal O}(z_{12})\right\},\cr
  F_2(z_i) = & (z_{12}z_{34}z_{13}z_{24})^{-\Delta}
  \left\{1+{\cal O}(z_{12})\right\},\cr
  F_3(z_i) = & (z_{12}z_{34})^{-2\Delta}
  \left\{1+{\cal O}(z_{12})\right\}.\cr}
  \eqn\Iiiia
  $$
On the other hand, $G(z_i)$ could equally
well have been evaluated in the limit as $z_2 \rightarrow z_3$. We would
have then found a new expansion in terms of conformal blocks:
  $$
  G(z_i) = F^\prime_3 (z_i) +
  {(d_{333})}^2 F^\prime_2 (z_i) +
  {(d_{335})}^2 F^\prime_1 (z_i).
  \eqn\Iiv
  $$
Since the conformal blocks form a complete basis of
$SL(2,\CC)$-invariant functions for $G$, we can write
  $$
  F_k (z_i) = {\alpha}_{kj} F^\prime_j (z_i)
  \eqn\Iv
  $$
for some matrix ${\alpha}_{kj}$ (the `fusion matrix').  For a
general four-point conformal block with fields labelled by
$a$, $b$, $c$, and $d$, the fusion matrix can be represented
diagrammatically as in Fig.~\fone.

Denote by $A_k$ the vector of structure constants
$A_k=\{(d_{335})^2,(d_{333})^2,1\}$ for $k=1,2,3$.
Demanding that \Iiii\ equal \Iiv\ gives the condition that $A_j$
must be an eigenvector of the fusion matrix with eigenvalue one:
  $$
  A_j \alpha_{jk} = A_k.
  \eqn\Ivi
  $$
In the case we are considering, $\alpha$ is a $3 \times 3$ matrix
$(j, k ~ \in ~ \{1,2,3\})$.  Expressions for the fusion matrices
for the minimal unitary series are collected in Appendix A, where
we find that, up to a similarity transformation that does not affect
the eigenvalues,
  $$
 \eqalign{
  \alpha_{jk}~~\sim~~ & \tilde{\alpha}^{(3)}_{jk}(-2,-2,-2;\rho+1)
  \cr & \cr =~~ & \left( \matrix{
    {\displaystyle {{x^4} \over {(1+x^4)(1+x^2+x^4)}}}
     & {\displaystyle {{-x^2} \over {1+x^4}}}
     & {\displaystyle {{x^2} \over {1+x^2+x^4}}} \cr
     {\scriptstyle {}} & {\scriptstyle {}} & {\scriptstyle {}} \cr
   {\displaystyle -{{1+x^2+x^4+x^6+x^8} \over {(1+x^4)(1+x^2+x^4)}}}
     & {\displaystyle {{1-x^2+x^4} \over {1+x^4}}}
     & {\displaystyle {{x^2} \over {1+x^2+x^4}}} \cr
     {\scriptstyle {}} & {\scriptstyle {}} & {\scriptstyle {}} \cr
   {\displaystyle {{1+x^2+x^4+x^6+x^8} \over {x^2(1+x^2+x^4)}}}
     & {\displaystyle 1 }
     & {\displaystyle {{x^2} \over {1+x^2+x^4}}}\cr }\right) , \cr }
  \eqn\IIxxiii
  $$
where $\rho = {1\over{K+2}}$ and $x={\rm exp}(i\pi\rho)$.
Since $\sum_k \alpha_{jk}\alpha_{kl}= \delta_{jl}$, the
eigenvalues of $\alpha$ must all be
$\pm1$.  It is easy to check that $\alpha$ has a double eigenvalue $+1$.
Thus, there is a one parameter family of solutions to \Ivi.

Is there some other condition on $A_j$ which picks out a single
solution?  A standard strategy is to impose locality (monodromy
invariance) on the correlation functions of our theory.  However,
the chiral $\phi_{3,1}$ fields
have fractional spins which create cuts in their OPEs.
We should expect correlation functions of these fields to be
multivalued, and for the locality condition to break down.
We can see this explicitly by computing the `braiding matrix'
for the conformal blocks of the $\phi_{3,1}$ four-point function.

Braiding comes into play when we interchange two fields in a correlation
function along some contour.  The locality condition states that the
resulting correlation function is equal to the same correlation function
with the arguments in the correct radial ordering:
  $$
  \langle \ldots \phi (z) \psi (w) \ldots \rangle =
  \langle \ldots \psi (w) \phi (z) \ldots \rangle
  \eqn\Ivii
  $$
for $|w| > |z|$, where the correlator on the left hand side means the
analytic continuation of the function $\langle \ldots \phi (w) \psi (z)
\ldots \rangle$ along a contour that interchanges $z$ and $w$.
If we perform this analytic continuation on the conformal blocks
of the four-point function, we obtain a new set of blocks
$\{F_k^{\prime\prime}(z_i)\}$.  These blocks are linearly related to
the original ones through the braiding matrix $\beta^+_{kj}$:\foot{The
superscript on $\beta^+$ refers to the particular choice for the
path along which the conformal blocks have been continued.
This choice is reflected in the sense of the over- or under-crossing
of the legs in Fig.~\ftwo.}
  $$
  F_k(z_i) = \beta^+_{kj} F_j^{\prime \prime}(z_i).
  \eqn\Iviii
  $$
The action of the braiding matrix can be represented diagrammatically
on four-point blocks as in Fig.~\ftwo.

The requirement of locality on four-point functions says that
$G(z_i) = A_k F_k (z_i) = A_k F^{\prime \prime}_k (z_i)$,
which implies that $A_j$ is an eigenvector of $\beta^+$ with
eigenvalue one:
  $$
  A_j \beta^+_{jk} = A_k.
  \eqn\Ix
  $$
In the case we are considering (a four-point function of the $\phi_{3,1}$
field) $\beta^+$ is a $3 \times 3$ matrix. So if it had a double
eigenvalue one, then generically there would be exactly one solution
of both \Ivi\ (associativity or fusion) and \Ix\ (locality or braiding).
However, in Appendix A we derive that (up to a similarity transformation)
  $$
  \beta^+_{jk}\sim \tilde{\beta}^{(3)+}_{jk}(-2,-2,-2;\rho+1)
  = (-1)^{j+k}x^{-j^2-k^2+7j+7k-16}\alpha_{jk}.
  \eqn\Ixa
  $$
It is easily checked that $\beta^+_{jk}$ does not, in general, have a
double eigenvalue one, and that there is {\it no} solution to both \Ivi\
and \Ix.  Therefore,
in order to consider fractional spin
chiral algebras we must drop the locality
assumption.  On general grounds we expect fractional spin fields
in two dimensions to satisfy more complicated braiding relations
than the usual Bose or Fermi statistics.  The braiding matrix
$\beta^+$ in \Ixa\ clearly shows that $\phi_{3,1}$ obeys a non-Abelian
braiding relation.

We are still left with a continuous family of possible structure
constants satisfying the associativity condition \Ivi.  In fact, this
extra freedom arises because we have not fully specified our fractional
supersymmetry chiral algebra.  The unitary model fusion rules implied
that the $\phi_{5,1}$ chiral field enters in the OPE \Iii.
Thus, we should also consider the OPEs of $\phi_{5,1}$ with $\phi_{3,1}$ and
with itself, typically generating yet more chiral fields.

Let us illustrate this with the $K=4$ minimal model.  In this case only the
$\phi_{5,1}$ field enters, and the minimal model fusion rules allow only the
following couplings:
  $$\eqalign{
  \phi_{3,1}\phi_{3,1}\sim & [1]+d_{333}[\phi_{3,1}]+d_{335}[\phi_{5,1}],\cr
  \phi_{3,1}\phi_{5,1}\sim & d_{335}[\phi_{3,1}],\cr
  \phi_{5,1}\phi_{5,1}\sim & d_{551}[1].\cr}
  \eqn\IIxiiia
  $$
We choose the normalization $d_{551}=1$, if it is not zero, by
changing the normalization of the $\phi_{5,1}$ field.  We have already
derived the associativity constraint \Ivi\ coming from the four-point
function of $\phi_{3,1}$ fields.  It turns out that the only other
constraint comes from the four-point function of two $\phi_{3,1}$
fields with two $\phi_{5,1}$ fields, as illustrated in Fig.~\fthree.
This translates to
  $$
  d_{551}=\alpha'(d_{335})^2
  \eqn\IIxiiib
  $$
where $\alpha'=\alpha^{(3)}_{33}(-2,-2,-4;\rho+1)$, in the notation
of Appendix A. (Note that even though $\alpha^{(3)}_{jk}(-2, -2,-4)$ is
generally a $3\times3$ matrix, when $K=4$ the fusion
rules truncate it to the one element $\alpha'$.)  Using the
formulas of Appendix A, it is a matter of simple algebra to
determine the solution to \Ivi\ and \IIxiiib\ to be $d_{333}=0$
and $d_{335}=1/\sqrt{\alpha'}$ ($\neq 0$).  Note, however, that there is a
second solution to \Ivi\ and \IIxiiib\ in which $d_{335}=d_{551}=0$,
and $d_{333}$ is non-zero.  The fact that $d_{551}$ vanishes in this
solution means that the $\phi_{5,1}$ field decouples entirely from
the chiral algebra.

Of these two solutions, only the second one ({\it i.e.}~ $d_{333}
\neq0$, $d_{335}=0$) persists for larger
values of $K$.  For $K\geq 5$, more couplings and more fields
enter into the fusion rules, but by checking a few cases
it is easy to see that the number of associativity
constraints increases faster than the number of couplings.
Thus generically there are
{\it no} solutions to the associativity constraints in which
none of the fields allowed by the fusion rules decouple.
(One can easily check the $K=5$ constraints explicitly to see
that no coincidences reduce the number of constraints to allow
such a solution.)

On the other hand, there clearly is a general solution (for
all $K\geq 3$) in which all the fields except $1$ and $\phi_{3,1}$
decouple from the chiral algebra.  Thus we will {\it define}
our chiral algebra to be the extended Virasoro algebra in which only the
$\phi_{3,1}$ field couples.  Then the only associativity constraint is
\Ivi\ coming from the $\phi_{3,1}$ four-point function,
subject to the condition $d_{335}=0$, or equivalently,
  $$
  A_1 = 0.
  \eqn\Ixi
  $$
This is the closure condition.
Because the fusion matrix $\alpha_{jk}$ in \Ivi\ has a double
eigenvalue $+1$, there is a unique solution to \Ivi\ subject to \Ixi.
Using the explicit formula \IIxxiii\ for the fusion matrix
(where now we must include the similarity factors, as described
in Appendix A), and the normalization of the $\phi_{3,1}$ field,
$A_3=1$, it is a matter of straightforward algebra to deduce
  $$
  (d_{333})^2=-{{4(K+6)^2}\over{3(K+4)(K+5)}}
  {{\sin^2(\pi\rho)\sin^2(4\pi\rho)}\over{\sin^3(2\pi\rho)\sin(3\pi\rho)}}
  {{\Gamma^3(\rho)\Gamma^2(4\rho)}\over{\Gamma(3\rho)\Gamma^4(2\rho)}},
  \eqn\IIxxv
  $$
where $\rho = {1\over{K+2}}$.  This, then, is the $\phi_{3,1}$ $\phi_{3,1}$
coupling in the chiral algebra in which the $\phi_{5,1}$ field does not
appear.

Before proceeding we should note that although \IIxxv\ gives the correct
result for $K=2$ and $K=3$, special features appear in these cases.
Considering the minimal model primary fields $\phi_{n,1}$ where
$1\leq n\leq K+1$, we see that $\phi_{5,1}$ does not exist for $K\leq 3$.
Therefore, for $K=3$, there are only two conformal blocks for the
$\phi_{3,1}$ four-point function and the $\alpha$ matrix \IIxxiii\ should
really be a $2\times 2$ matrix.  The correct $\alpha$ matrix is easily found,
however, by truncating the first row and column of \IIxxiii. This, in essence,
is what we did by imposing the closure condition $A_1=0$ \Ixi.
For $K=2,$ life is even simpler. From the fusion rule \IIxxva,
we see that the $\phi_{3,1}~\phi_{3,1}$ OPE closes only on the identity.
Therefore, we know {\it a priori}\/ that $d_{333}=0$ for $K=2$.  Indeed,
\IIxxv\ gives this result. Strictly speaking, the derivation given above
breaks down since the $\alpha$ matrix \IIxxiii\ is ill-defined: some of
the entries are infinite. This is due to the degeneracies among the
conformal blocks, and is a general feature whenever $K$ is not sufficiently
large.

As was explained in the Introduction, the minimal model with
$c=1-6/(K+2)(K+3)$ is a representation of the
fractional supersymmetry chiral algebra ${\cal A}^{(K)}$ \Zi,
with $\phi_{3,1}$ playing the role of the \JK\ current.  Thus
$d_{333}$ is the structure constant for the minimal model value of the
central charge, {\it i.e.}~$\lambda_K(c_{\rm min})=d_{333}(K)$.

To avoid any confusion, we should note that the value for the
structure constant $d_{333}$ \IIxxv\ derived
above is not the same as that found by Dotsenko and Fateev and others
[\DotFat,\FFK].  The reason is that they compute the structure constants
for a {\it non-chiral} theory.  In particular, they consider four-point
correlation functions of left-right symmetric combinations of chiral
fields and demand associativity of the operator product algebra, and
monodromy invariance (locality) of the correlator. For example, in the
unitary model, to calculate $\hat d_{333}$ they consider the four-point
function $G(z_i,\bar z_i) = \langle\phi_{3,1}(z_1,\bar z_1)
\phi_{3,1}(z_2,\bar z_2) \phi_{3,1}(z_3,\bar z_3)
\phi_{3,1}(z_4,\bar z_4)\rangle$
where the $\phi_{3,1}$ OPE is
  $$\eqalign{
  \phi_{3,1}(z,\bar z)\phi_{3,1}(w,\bar w)= & |z-w|^{-4\Delta}[1] +
  |z-w|^{-2\Delta}|\hat d_{333}|^2 [\phi_{3,1}(w,\bar w)] \cr
  & + |z-w|^{2\Delta+2}|\hat d_{335}|^2 [\phi_{5,1}(w,\bar w)]. \cr}
  \eqn\Ixiii
  $$
It is crucial to note that the structure constants that appear in \Ixiii\
are not the same ones that appear in the chiral OPE \Iii.
Since there are no ``off-diagonal'' combinations of holomorphic and
antiholomorphic vertex operators appearing in \Ixiii, it cannot be simply
the product of left- and right-moving chiral OPEs.  The two ways of fusing
fields in $G$ give the relation between conformal blocks
$\hat A_k|F_k (z_i)|^2 = \hat A_k|F^\prime_k (z_i)|^2$  which in turn implies
  $$
  \sum_l \hat A_l \alpha_{lj} \bar\alpha_{lk} =
   \hat A_j \delta_{jk}.
  \eqn\Ixiiia
  $$
This determines the structure constants $\hat d_{333}$
and $\hat d_{335}$ uniquely.  In particular, $\hat d_{333}\neq d_{333}$
and $\hat d_{335}\neq 0$, so it is clear that this non-chiral
solution to the associativity constraints is different from
the chiral solutions described above.

The fact that the fields in the non-chiral model
only appear in left-right symmetric combinations ensures that
once \Ixiiia\ is satisfied, so is invariance of the correlator
under braiding transformations.  The point is that by considering
only left-right symmetric combinations, Dotsenko and Fateev
restrict themselves to integer- (actually zero-) spin fields,
which have trivial braiding properties.  We will discuss the braiding
(or locality) properties of our chiral solutions in Section 5.


\chapter{Generalized Feigin-Fuchs Realization of \AK}

It was noticed some time ago in ref. [\Kastor,\Bagger,\Rava]
that conformal field theories with \AK\ \Zi\ as their chiral
algebra can be constructed by a generalized Feigin-Fuchs technique.
One `bosonizes' the $SU(2)_K$ WZW theory with a boson
and a $Z_K$-parafermion, and then introduces a background charge for
the boson. For special values of the
background charge, unitary representations of ${\cal A}^{(K)}$ are found,
corresponding to the coset models $SU(2)_K \otimes SU(2)_L / SU(2)_{K+L}$
with $L$ varying over the positive integers for various background charges.

In the generalized Feigin-Fuchs construction,
the screening operators are $V_+=\psi_1{\rm exp}(i\sqrt2\alpha_+\varphi)$
and $V_-=\psi_1^\dagger{\rm exp}(i\sqrt2\alpha_-\varphi),$ where $\psi_1$
and $\psi_1^\dagger$ are $Z_K$-parafermion currents [\ZFPara].
$\varphi$ is a scalar field with background charge with
  $$ \matrix{
  T_\varphi(z)=-{1\over2}{(\partial\varphi)}^2+i\sqrt2
  \alpha_0\partial^2\varphi,
  & c_\varphi = 1-24\alpha^2_0, \cr
  \langle \varphi (z) \varphi (w) \rangle = - {\rm ln}(z-w),
  & \alpha _\pm = \alpha _0 \pm \sqrt{\alpha^2_0+{1\over K}}.\cr}
  \eqn\IIIiii
  $$
The energy-momentum tensor of the ${\cal A}^{(K)}$ algebra is then
$T = T_\varphi + T_{Z_K},$ and the total central charge is
  $$
  c=c_\varphi+c(Z_K)=(1-24\alpha^2_0)+{{2(K-1)}\over{K+2}}.
  \eqn\IIIcent
  $$
Recall that the current \crnt z has dimension $\Delta =(K+4)/(K+2)$.
Searching the $Z_K$-parafermion spectrum for fields of the appropriate
dimension reveals only four potential candidates:
  $$ \matrix{
  \epsilon_1 = A^\dagger_{-{1\over K}} \sigma_2, &
  \epsilon_1^\dagger = A_{-{1\over K}} \sigma_2^\dagger, \cr
  \hat \epsilon_1 = A^\dagger_{-{1\over K}-1} \sigma_2, &
  \hat \epsilon_1^\dagger = A_{-{1\over K}-1} \sigma_2^\dagger, \cr}
  \eqn\IViii
  $$
where $A$ and $A^\dagger$ are the modes of the $\psi_1$ and
$\psi_1^\dagger$
parafermion currents and $\sigma_2$ and
$\sigma^\dagger_2 \equiv \sigma_{K-2}$
are $Z_K$-parafermion highest weight operators (spin fields).
The definition \IViii\ of the $\epsilon$-fields in terms of parafermion
current modes is equivalent to the following OPE:
  $$
  \psi_1^\dagger(z)\sigma_2(w)=(z-w)^{{2\over K}-1}\left\{
  \epsilon_1(w)+(z-w)\hat\epsilon_1(w)+\ldots\right\},
  \eqn\IViiiaa
  $$
as well as its daggered version.  The ``energy'' operators $\epsilon_1,
\epsilon^\dagger_1$ have dimension $\Delta-1$ and their parafermion
descendants $\hat\epsilon_1, \hat\epsilon^\dagger_1$ have dimension
$\Delta$.
One can show [\ALT], using parafermion current algebra identities,
that these four fields are not independent, but are related by
 $$
 \epsilon_1^\dagger = \epsilon_1,\qquad\qquad
 \hat\epsilon_1^\dagger+\hat\epsilon_1=
 \left({K+2}\over2\right)\partial\epsilon_1.
 \eqn\IViiia
 $$
Thus there are only three independent fields of the correct dimension to
form $J^{(K)}$: $\epsilon_1\partial\varphi$, $\partial\epsilon_1$, and
$\hat\epsilon_1$.

To construct $J^{(K)}$ we search for the combination of the above fields
that commutes with the screening charges $V_\pm$ [\Kastor,\Bagger,\Rava].
Using the $\psi_1, \psi_1^\dagger$ OPEs with
$\epsilon_1, \hat \epsilon_1:$
  $$\eqalign{
  \psi_1(z) \epsilon_1(w) = & \left({2 \over K}\right)
  {{\sigma_2(w)} \over {(z-w)}} +
  \left({{K+2} \over K}\right) \partial \sigma_2(w) + \ldots \cr
  \psi_1(z) \hat \epsilon_1(w) = & \left({{K^2+2K-4}
  \over {K^2}}\right) {{\sigma_2(w)}
  \over {{(z-w)}^2}} - \left({{2K+4} \over {K^2}}\right)
  {{\partial \sigma_2(w)} \over
  {(z-w)}} + \ldots \cr}
  \eqn\IVv
  $$
as well as the daggered versions of these
relations, the condition that \JK\
commutes with $V_{\pm}$ determines the
relative normalizations of the possible
terms [\ALT,\BerLe]
  $$
  J^{(K)}(z) = \sqrt{{K\Delta}\over{4c}} \left\{ [\sqrt 2 \epsilon_1
  \partial \varphi - i \alpha_0 (K+2) \partial \epsilon_1] +
  {{i K (\alpha_+ - \alpha_-)}\over{K+4}} [\hat \epsilon_1 -
  \hat \epsilon^\dagger_1]\right\},
  \eqn\IVvi
  $$
where the overall normalization has been chosen to match \Zi. The terms in
square brackets are Virasoro primary combinations.

Now that we have an explicit expression for \crnt{z}, it
is possible to calculate the chiral algebra (\AK) that it satisfies.
Because \IVvi\ is Virasoro primary, the part of \AK\ involving
the energy-momentum tensor is standard:
  $$\eqalign{
  T(z)T(w)= & {c\over 2}(z-w)^{-4}\left\{1+{4\over c}(z-w)^2T(w)
   +\ldots\right\},\cr
  T(z)J^{(K)}(w)= & \Delta (z-w)^{-2}\left\{J^{(K)}(w)
   +{1\over \Delta}(z-w)
   \partial J^{(K)}(w)+\ldots\right\},\cr}
  \eqn\IVZia
  $$
where the dots denote the Virasoro descendants appearing with higher integer
powers of $(z-w)$.  The \JK\JK\ OPE, of the form
  $$\eqalign{
  J^{(K)}(z)J^{(K)}(w)= & (z - w)^{-2\Delta}\left\{1+
   {{2\Delta}\over c}(z-w)^2
   T(w)+\ldots\right\}\cr
    & +\lambda_K(c)(z - w)^{-\Delta}\left\{J^{(K)}(w)+{1\over 2}(z-w)
     \partial J^{(K)}(w)+\ldots\right\}+\ldots,\cr}
  \eqn\IVZib
  $$
would at first sight seem to be easy to calculate:  all we have to do is
take the OPE of \IVvi\ with itself. However, when we do that we find that
$\lambda_K (c)$ is only determined in terms of the $SU(2)_K$ structure
constant $c_{111}$, which governs the fusing of two chiral spin-$1$ primary
fields to give a spin-$1$ field.

To compute the OPE of \crnt z with itself as it is
given in \IVvi, we need the leading terms of the OPEs
of the parafermion fields $\epsilon_1 (z)$ and
$\eta (z) \equiv \hat \epsilon_1 (z) - \hat \epsilon^\dagger_1 (z)$.
(Note that when $K=2$, $\hat\epsilon_1(z)=\hat\epsilon^\dagger_1(z)$,
so $\eta(z)$ is absent.)
In Appendix B we calculate these OPEs using the definition of the
$Z_K$-parafermion theory in terms of the $SU(2)_K$ WZW
theory, plus the $SU(2)_K$ Ward identities.  The results
for the leading terms of interest to us are
  $$\eqalign{
  \epsilon_1(z)\epsilon_1(0)= & -c_{110}{2\over K}z^{-2\Delta+2}
  +c_{111}{{2\sqrt{K}}\over{(K-2)(K+4)}}z^{-\Delta+2}\eta(0)+\ldots\cr
  \eta (z)\eta(0)= & -c_{110}{{2(K-2)(K+4)}\over{K^2}}z^{-2\Delta}
  -c_{111}{{3(K-4)}\over{\sqrt{K}(K-2)}}z^{-\Delta}\eta(0)+\ldots\cr
  \epsilon_1(z)\eta(0)= & \eta(z)\epsilon_1(0)=
   c_{110}{\cal O}(z^{-2\Delta+2})
  +c_{111}{{2}\over{\sqrt K}} z^{-\Delta} \epsilon_1 (0) + \ldots \cr}
  \eqn\IVvii
  $$
Here $c_{110}$ is the $SU(2)_K$ structure constant for
two $SU(2)$ spin-$1$ primaries
to fuse to the identity. It can be set
to one, fixing the normalization of the
spin-$1$ fields; we will do so henceforth.
$c_{111}$ is the chiral $SU(2)_K$
structure constant mentioned before.
By ${\cal O}(z^{-2\Delta+2})$ in the
last equation, we mean only that the
leading term proportional to
$c_{110}z^{-2\Delta+1}$ vanishes.

Using these OPEs and the explicit expression \IVvi\ for \JK, we can
calculate the leading terms in the $J^{(K)}J^{(K)}$ OPE.  In particular,
the coefficient of the identity in
\IVZib\ is determined by the $c_{110}$ pieces
in the $(\epsilon_1\partial\varphi)(\epsilon_1\partial\varphi)$,
$\partial\epsilon_1\partial\epsilon_1$, and $\eta\eta$ OPEs.  Potential
contributions from $\partial\epsilon_1\eta$ and $\eta\partial\epsilon_1$
OPEs do not contribute since the $c_{110}$ term of the $\epsilon_1\eta$
OPE is of order ${\cal O}(z^{-2\Delta+2})$.  The normalization
of \JK\ in \IVvi\ was actually determined by demanding that the
\JK\JK\ OPE close on the identity with coefficient one.  The coefficient
\lK\ of \JK\ in \IVZib\ can be extracted by computing the
coefficient of the $\epsilon_1\partial\varphi$ terms.  The $c_{111}$
pieces of the $\epsilon_1\eta$,
$\eta\epsilon_1$, $\epsilon_1\partial\epsilon_1$
and $\partial\epsilon_1\epsilon_1$ OPEs contribute, giving
  $$
  \lambda^2_K (c) = {{2 K^2 (c_{111})^2}\over{3(K+4)(K+2)}}
  \left[{{3 (K+4)^2}\over{K(K+2)}}{1\over c} - 1\right]
  \eqn\IVviii
  $$
where we have used the relation between the central charge and
background charge \IIIcent.  This argument evaluates \lK\ by computing
only the coefficient of the $\epsilon_1\partial\varphi$ term in
\JK.  A similar computation using the $c_{111}$ pieces of the
$\epsilon_1\epsilon_1$ and $\eta\eta$ OPEs
shows that the \JK\JK\ OPE indeed has the form \IVZib, closing
precisely on \JK.  Other terms, not included in (3.10), indicate that
one other chiral primary field, proportional to parafermion
descendants of the $\sigma_4$ spin field, may enter in the \JK\JK\
OPE.  The structure constant for this additional field
is proportional to $c_{112}$, the chiral $SU(2)_K$ structure
constant for two spin-$1$ fields to fuse to a spin-$2$ field.
Since we have defined the fractional supersymmetry algebras
to be those in which no fields other than $T(z)$ and \crnt{z}
(and their Virasoro descendants) appear, we must demand that
$c_{112}=0$.  This is the closure condition, discussed in the
last section.  In the next section we will show that it is
compatible with the associativity constraints of the chiral
$SU(2)_K$ WZW model.

If we fix the background charge in \IIIcent\ so that the central
charge takes on the minimal unitary
model values $c_{\rm min}=1-6/(K+2)(K+3)$
we know from the discussion in the previous section
that $\lambda_K(c_{\rm min})=d_{333}$ (where we have chosen to
normalize the $\phi_{3,1}$ field in the same way that $J^{(K)}$ is
normalized, {\it i.e.}~$d_{331}=1$). Comparing to \IVviii\
gives a relation between the structure constants:
  $$
  (d_{333})^2 = {4\over 3} (c_{111})^2 {{(K+6)^2}\over
  {(K+4)(K+5)}}.
  \eqn\IVxi
  $$
Using the value for $d_{333}$ \IIxxv\ computed by associativity and
closure, we obtain the explicit expression for $c_{111}$ \Zia\
given in the Introduction.

{\it A priori} the $SU(2)_K$ structure constant $c_{111}$ could depend
on the central charge, thereby invalidating the argument leading
to \IVxi.  Indeed, $c_{111}$ should be thought of as a \sc\ in
the coset model representation of \AK, and therefore it should be
determined by requiring associativity in that model.  However, because
of the simple way in which the Feigin-Fuchs boson $\varphi$ enters into the
expression \IVvi\ for \JK, $c_{111}$ can be identified as the \sc\
determined by associativity of the chiral $SU(2)_K$ theory.  In
particular, correlators involving \JK\ can be broken up into sums of
products of $Z_K$-parafermion correlators and Feigin-Fuchs boson correlators.
Since the latter only involve the boson ``current'' $\partial\varphi$,
they are independent of the background charge $\alpha_0$, and the
chiral $Z_K$-parafermion correlators must satisfy associativity
by themselves.  We will show in the next section that, just as in
the minimal model case examined in the last section, there is
only one solution to the associativity and closure conditions for
the chiral $SU(2)_K$ WZW model and thus only one for the
chiral $Z_K$-parafermion model since it is simply the WZW
model stripped of a free boson.  Therefore, for given $K$,
$c_{111}$ should be independent of the central charge
of the generalized Feigin-Fuchs realization of \AK.


\chapter{Associativity Constraints in Chiral $SU(2)_K$ WZW Models}

Setting the background charge to zero in the generalized Feigin-Fuchs
representation of the last section corresponds to taking the
$L\rightarrow\infty$ limit of the $SU(2)_K\otimes SU(2)_L/SU(2)_{K+L}$
coset model.  In this limit we expect to recover the $SU(2)_K$
WZW theory.  Now, the $SU(2)_K$ WZW model has been solved [\Dots]
by using the Wakimoto representation [\Waki] to represent the spin-$j$
primary fields $\Phi_m^{(j)}$ in terms of free fields with background
charge, allowing the derivation of integral expressions for the
conformal blocks.  It is found [\Dots] that the fusion matrices
of the $SU(2)_K$ spin-$j$ primary fields are the same up to similarity
transformations as those of the $\phi_{2j+1,1}$ primaries of the
$K$th minimal unitary model.  The similarity transformations
depend on the normalization constants of the conformal blocks, but do
not affect the eigenvalues of the fusion matrices.  This means that the
analysis of associativity constraints presented for the minimal model
in Section 2 works through in precisely the same way for the WZW theory
since it only depends on the number of $+1$ eigenvalues of the fusion
matrix.  In particular, we are still assured of a solution to the
associativity conditions if we demand that only the energy momentum
tensor and the spin-$1$ field (corresponding to $\phi_{3,1}$) appear,
all other fields decoupling from our chiral algebra.
In this case $c_{111}$
is determined by the analog of eqn.~\Ivi, where $\alpha_{jk}$ is now
the $SU(2)_K$ fusion matrix and $A_k=\{{1\over 6}(c_{112})^2,
 {1\over 2}(c_{111})^2,1\}$ is the vector of $SU(2)_K$ structure constants.
The factors of ${1\over 6}$ and ${1\over 2}$ in the vector $A_k$ are
standard $SU(2)$ group factors explained in Appendix B, (B12)-(B13).
The closure condition,
corresponding to \Ixi, is $c_{112}=0$, and tells us that the spin-$2$
fields decouple.  Using the results for the normalization constants for
the $SU(2)_K$ conformal blocks [\Dots], summarized in Appendix C, we find the
expression \Zia\ for $c_{111}$ given in the Introduction.

It is important to note that this way of determining $c_{111}$
is independent of the argument used in the last section, where
it was fixed by comparing to the results of the minimal model
associativity argument of Section~2.  The fact that both
give the same result is a nontrivial check of the calculation
of $\lambda_K(c)$ using the generalized Feigin-Fuchs representation
of Section~3.

We have made the vague identification of the WZW spin-$1$
primary with the \JK\ current in the last two paragraphs.
Actually, \JK\ is a Kac-Moody descendant of the WZW spin-$1$
primary field.  To see this, take $\alpha_0=0$ in the generalized
Feigin-Fuchs model of the previous section.  The limit of
\IVvi\ gives the fractional current in terms of parafermion
fields and a free boson:
  $$
  J^{(K)}(z)=\sqrt{{K+4}\over{6}}\left\{\epsilon_1(z)\partial\varphi(z)
  +{{i\sqrt{2K}}\over{(K+4)}}\eta(z)\right\}.
  \eqn\Vi
  $$
Using the standard connection between $Z_K$-parafermions plus a free
boson and the $SU(2)_K$ WZW theory (reviewed in Appendix B) we can
reexpress \Vi\ as
  $$
  J^{(K)}(z)={i\over{\sqrt{3(K+4)}}}\left(J^-_{-1}\Phi_+^{(1)}(z)
  -\sqrt{2}J^0_{-1}\Phi_0^{(1)}(z)-J^+_{-1}\Phi_-^{(1)}(z)\right),
  \eqn\Vii
  $$
where $J^a_{-1}$ are Kac-Moody current modes and $\Phi_m^{(1)}(z)$ is
the WZW spin-$1$ primary field with dimension $2/(K+2)$. (This expression is
proportional to $q_{ab}J^a_{-1}\Phi_{(1)}^b$, the form for $J^{(K)}$
mentioned in Table~\tone.)

The associativity condition we are really interested in involves the
\JK\ current, and not the spin-$1$ fields.  To fix the \JK\JK\ \sc\
in this theory requires knowledge of the fusion matrix for the
four-point function of $J^{(K)}$ currents.  It is not obvious, {\it a
priori}, that this fusion matrix is the same as that for the spin-$1$
fields derived in [\Dots] (and summarized in Appendix C).  However,
it turns out that the fusion matrix for the \JK\ field differs from
that of the spin-$1$ field only by a similarity transformation.
This is in line with the observation that the fusion matrix
reflects the structure of cuts in the OPEs of the chiral fields which in
turn depends only on the fractional part of the dimensions of those fields.
Since the dimension of $J^{(K)}$ differs from that of the WZW spin-$1$ field
by an integer (one), we expect the fusion matrix to be the same up to a
similarity transformation.  Of course, the actual solution for the structure
constants depends on that similarity transformation.

In the remainder of this section, we will outline the calculation of
the \JK\ four-point conformal block fusion matrix in the chiral $SU(2)_K$
WZW model.  We will show that it implies a value for \lK\ in agreement
with the one found by our previous arguments.

We follow ref.~[\Dots] in which Dotsenko uses the Wakimoto representation
[\Waki] to solve the $SU(2)_K$ WZW model.  We introduce a scalar field
with background charge $\alpha_0$ and a $c=2$ system, $\omega$
and $\omega^+$, of dimension 0 and 1 respectively, satisfying
  $$
  \langle\omega(z)\omega^+(w)\rangle=
  -\langle\omega^+(z)\omega(w)\rangle={i\over{z-w}}.
  \eqn\Viv
  $$
The combined boson plus $(\omega,\omega^+)$ system gives a realization
of the $SU(2)_K$ WZW model.  Since the total central charge of this
system is $c=3-24\alpha_0^2$, the level must be given by
$K+2=1/4\alpha_0^2$.  The screening momenta are $\alpha_+=2\alpha_0$ and
$\alpha_-=0$.  The screening charge $Q_+=\oint\!dz\,
\omega^+(z)V_{\alpha_+}(z)$, where $V_\alpha(z)={\rm exp}
\{i\sqrt2\alpha\varphi(z)\}$, commutes with the Kac-Moody currents
given by [\Waki]
  $$\eqalign{
  J^+= &\ \omega^+,\cr
  iJ^0= &\ \omega\omega^++{1\over{2\sqrt{2}\alpha_0}}\partial\varphi,\cr
  J^-= &\ \omega^2\omega^++iK\partial\omega+
  {1\over\sqrt{2}\alpha_0}\omega\partial\varphi.\cr}
  \eqn\Vvi
  $$
The spin-$j$ primary fields are
expressed in terms of binomial coefficients and
the free fields as (in the normalization of Appendix B)
  $$
  \Phi_m^{(j)}(z)=i^{j-m}\left(\matrix{2j\cr j+m\cr}\right)^{1/2}
  (\omega(z))^{j-m}V_{\alpha_j}(z),
  \eqn\Vvii
  $$
where $\alpha_j=-2\alpha_0j$.

{}From these formulas we can immediately
write down the expression for $J^{(K)}$
in terms of free fields
  $$
  J^{(K)}=-\sqrt{{{K+4}\over 3}}\,\partial\omega\,V_{-\alpha_+}.
  \eqn\Vviii
  $$
In order to evaluate the $J^{(K)}$ four-point correlation function
we must construct the Feigin-Fuchs conjugate current $\tilde{J}^{(K)}.$
A field and its Feigin-Fuchs conjugate have the same conformal and
$SU(2)_K$ properties; both are necessary in order to cancel the
background charges and construct non-zero correlation functions in
the Wakimoto representation.  The conjugate to the identity operator
is found to be [\Dots]
  $$
  \tilde{1}=(\omega^+)^{-K-1}V_{-\alpha_+(K+1)}.
  \eqn\Vviiia
  $$
This implies the neutrality conditions for correlation functions,
$N^--N^+=K+1$ and $\sum\alpha_i=-\alpha_+ (K+1)$, where $N^+$ and
$N^-$ are the number of $\omega^+$ and $\omega$ zero-modes, respectively,
and the $\alpha_i$ are the ``momenta'' of the $\varphi$ zero-modes
entering in the correlator. The construction of the fields conjugate
to the spin-$j$ primaries, $\tilde{\Phi}^{(j)}_m$, is outlined in [\Dots].
Using the spin-$1$ conjugates and \Vii\ it is straightforward to construct
the conjugate of \JK:
  $$\eqalign{
  \tilde{J}^{(K)}= & -\sqrt{{{K+4}\over 3}}
  \Biggl\{(\partial\omega)(\omega^+)
   -{{iK(K^2-1)}\over{2(K+4)}}(\omega^+)^{-1}(\partial^2\omega^+)\cr
  & +{iK(K^2-1)\over 2}(\omega^+)^{-2}(\partial\omega^+)^2
   -{{K(K-1)(2K+5)}\over{2\sqrt{2}\alpha_0(K+4)}}(\partial\varphi)
   (\omega^+)^{-1}(\partial\omega^+)\cr
  & +{{(K-1)(K-2)}\over{2\sqrt{2}\alpha_0(K+4)}}(\partial^2\varphi)
   -{{i(K^2-1)(K+2)}\over{(K+4)}}(\partial\varphi)^2\Biggr\}
   (\omega^+)^{-K}V_{-K\alpha_+}.\cr}
  \eqn\Vviiib
  $$

We can now write the \JK\ four-point correlator as
  $$
  G_J(z_i)=\langle J^{(K)}(z_1) J^{(K)}(z_2) J^{(K)}(z_3)
  \tilde J^{(K)}(z_4) Q_+ Q_+\rangle,
  \eqn\Vix
  $$
where the two insertions of the screening charge $Q_+$ are
necessary to satisfy the neutrality conditions mentioned
above.  Using the explicit expressions derived above
for \JK, $\tilde J^{(K)}$, and $Q_+$, $G_J(z_i)$ can
be written as a sum of products of $\omega$, $\omega^+$
and $\varphi$ free-field correlators (modulo the zero-mode
insertions required to satisfy the neutrality constraints).
However, the contours for the integrals in the definition
of the screening charges has not been specified in \Vix.
As in the usual Feigin-Fuchs procedure [\DotFat], there
are only three linearly independent choices of contours
for two screening charges in a four-point function.
Thus, evaluating the free-field correlation functions
we can write $G_J(z_i)$ in terms of three conformal blocks
  $$
  G_J(z_i)=(z_{13}z_{24})^{-2\Delta}[\eta(1-\eta)]^{2\over{K+2}}
  \sum^3_{k=1}A_k(N_k)^{-1}I_k(\eta),
  \eqn\Vx
  $$
where $z_{ij}=z_i-z_j$ and $\eta$ is the projective invariant
$\eta=z_{12}z_{34}/z_{13}z_{24}$.  The $A_k$ are the as yet
undetermined coefficients of the blocks, and the $N_k$'s are
normalization constants. The integral expressions for the
conformal blocks have the following form
  $$
  \left.\matrix{I_1(\eta)\!\!\!&\!\!=\!\!\!&\!\!\!
    \int^\infty_1\!du_1\int^{u_1}_1\!du_2\cr
  I_2(\eta)\!\!\!&\!\!=\!\!\!&\!\!\!
    \int^\infty_1\!du_1\int^{\eta}_0\!du_2\cr
  I_3(\eta)\!\!\!&\!\!=\!\!\!&\!\!\!
    \int^{\eta}_0\!du_1\int^{u_1}_0\!du_2\cr }\right\}
  [u_1u_2(u_1-\eta)(u_2-\eta)(u_1-1)(u_2-1)]^{-2{\rho}}
  (u_1-u_2)^{2{\rho}}R(u_1,u_2,\eta),
  \eqn\Vxb
  $$
where
  $$\eqalign{
  R(u_1,u_2,\eta)= & {1\over{u_1^2(u_2-\eta)^2}}+{1\over{u_1^2(u_2-1)^2}}
  +{1\over{(u_1-\eta)^2u_2^2}}\cr & +{1\over{(u_1-\eta)^2(u_2-1)^2}}
  +{1\over{(u_1-1)^2u_2^2}}+{1\over{(u_1-1)^2(u_2-\eta)^2}},\cr}
  \eqn\Vxc
  $$
and ${\rho}=1/(K+2)$. The normalization constant
$N_k$ is defined as the coefficient of the leading term in $\eta$ of
$I_k(\eta)$ as $\eta\rightarrow0$. Using the results in [\DotFat] we find
that
  $$\eqalign{
  N_1=&\left[{{-10\pi^2(K+2)^2(K+4)}\over{3(K+5)(K+6)}}\right]
  \left[{1\over{\sin(\pi\rho)\sin(2\pi\rho)}}\right]
  \left[{{\Gamma(5\rho)}\over{\Gamma^2(\rho)\Gamma(3\rho)}}\right]\cr
  N_2=&\left[{{-16\pi^2(K+2)^2}\over{(K+4)^2}}\right]
  \left[{{\sin(4\pi\rho)}\over{\sin^3(2\pi\rho)}}\right]
  \left[{{\Gamma^2(4\rho)}\over{\Gamma^4(2\rho)}}\right]\cr
  N_3=&\left[{{-3\pi^2(K+2)^2}\over{(K+4)}}\right]
  \left[{{\sin(3\pi\rho)}\over{\sin^2(\pi\rho)\sin(2\pi\rho)}}\right]
  \left[{{\Gamma(3\rho)}\over{\Gamma^3(\rho)}}\right].\cr}
  \eqn\Vxca
  $$

As $z_1\rightarrow z_2$, we find that $G_J(z_i)$ has the leading
order behavior,
  $$\eqalign{
  G_J(z_i)=
  &A_3(z_{12}z_{34})^{-2\Delta}
   \{1+{\cal O}(z_{12})\}
  +A_2(z_{12}z_{34}z_{13}z_{24})^{-\Delta}
   \{1+{\cal O}(z_{12})\}\cr
  &+A_1(z_{12}z_{34})^{\Delta-1}(z_{13}z_{24})^{-3\Delta+1}
   \{1+{\cal O}(z_{12})\}.\cr}
  \eqn\Vxcb
  $$
We are now able to make the identification, as in Section 2,
of the $A_i$'s with the structure constants of the ${\cal A}^{(K)}$
chiral algebra.  In particular if we set $A_3=1$ to properly
normalize the $J^{(K)}$ current, then $A_2=\lambda_K^2$.
The first term of \Vxcb\ implies that a field of dimension $3\Delta-1$
enters into the \JK\JK\ OPE (whereas the dimension of the
$\phi_{5,1}$ field that entered into the $\phi_{3,1}\phi_{3,1}$ OPE
in Section 2 was $3\Delta+1$).  In fact, the dimension $3\Delta-1$
field corresponds to a level 2 Kac-Moody descendant of the spin-2
primary field of dimension $3(\Delta-1)$.

The $I_k$ integrals are related under the change of variables
$\eta\rightarrow 1-\eta$ by the matrix $\tilde{\alpha}^J$:
  $$
  I_j(\eta)=\tilde{\alpha}^J_{jk}I_k(1-\eta).
  \eqn\Vxea
  $$
We calculate $\tilde{\alpha}^J$ using the contour deformation
techniques described in Appendix A. Briefly, we make the change of
variables $\eta\rightarrow 1-\eta$ and $u_i\rightarrow 1-u_i$
in the $I_k$ integrals \Vxb\ which changes the limits of integration but
leaves the integrand invariant.
We can then pull the contours of integration
back to their original positions and obtain \Vxea .
In actuality we do not need
to do all of this work, since the rational function $R$ is merely a
spectator in all of these manipulations; its effects are manifest only in the
normalization constants $N_k$.  The rest of the integrand is of the same form
as the integrands that appear in the minimal model conformal blocks (A4).
Therefore, in the notation of Appendix~A (A11), we find
  $$
  \tilde{\alpha}^J_{jk}=
  \tilde{\alpha}^{(3)}_{jk}(-2,-2,-2;\rho).
  \eqn\Vappa
  $$
This is the matrix that appeared (2.9) in our discussion of the
minimal model associativity condition in Section~2.  Since
${\rho}=\tilde\rho-1$ we have that ${x}=\exp(i\pi\rho)=-\tilde x$,
but since (2.9) is even in $x$, there is no difference between the
two matrices.

Examining \Vx\ we see that the fusion matrix for the conformal blocks
of the $J^{(K)}$ four-point functions is given by
$(N_j)^{-1}\tilde{\alpha}^J_{jk}(N_k)$.  We can now determine
$\lambda_K$ by requiring that the $A_k$'s satisfy the
associativity condition, given by the eigenvalue equation
  $$
  \sum_j A_j(N_j)^{-1}\tilde{\alpha}^J_{jk}(N_k)=A_k,
  \eqn\Vxe
  $$
and by imposing the closure condition $A_1=0.$ This gives the result
  $$
  \lambda^2_K(c_{SU(2)})={{16}\over{3(K+4)}}
  {{\sin^2(\pi\rho)\sin^2(4\pi\rho)}\over{\sin^3(2\pi\rho)\sin(3\pi\rho)}}
  {{\Gamma^3(\rho)\Gamma^2(4\rho)}\over{\Gamma(3\rho)\Gamma^4(2\rho)}},
  \eqn\Vxii
  $$
in agreement with \Zii\ and \Zia.


\chapter{Discussion}

In this section we briefly comment on a few interesting questions
associated with the \FS\ \ca s \AK, and with fractional spin \ca s in
general.  (a) We will first describe the braiding properties of the
\FS\ currents \JK.  (b) Next, we investigate the possible existence
of new \ca s differing from the \FS\ \ca s by having two or more
\JK\ currents, although still with the same dimensions as in the
\AK\ algebra.  (c) Finally, we address the possibility of constructing
in our framework non-local \ca s with a fractional spin current
of dimension different from that of \JK.

\noindent {\bf a.}~~~It was shown in Section~2
that the fractional spin current \JK\
does not satisfy simple Bose or Fermi statistics.  Instead, the
correlation functions of these currents transform according to
some more complicated representation of the braid group, given by
the braiding matrices $\beta^\pm_{jk}$ introduced earlier.
We can examine the properties of the four-point conformal block
braiding matrices by relating them to the fusion matrix and
``twisting'' matrices $\gamma^\pm$ [\MooSigh,\FFKI].  The latter
are just the diagonal matrices of phases of conformal blocks
as two operators which are ``contracted'' are exchanged.  More
concretely, consider two fields $\phi_1(z_1)$ and $\phi_2(z_2)$
in a block which is normalized as $z_1\rightarrow z_2$.  The
phases under interchange of these two fields are simply
determined by the powers of $(z_1-z_2)$ appearing in their OPE:
if this OPE contains a field $\phi_i(z_2)$ on the right hand side,
then the corresponding monodromy is
${\rm exp}\{\pm i\pi(\Delta_i-\Delta_1-\Delta_2)\}$.  So, in general
we have
  $$
  \gamma^\pm_j(a,b)={\rm e}^{\pm i\pi(\Delta_j-\Delta_a-\Delta_b)}.
  \eqn\Vgam
  $$

The relations between $\alpha$, $\beta^\pm$, and $\gamma^\pm$ that
we want are most easily described diagrammatically.  We adopt the
notation shown in Fig.~\ffour\ for the action of $\gamma$ on
conformal blocks (as well as the notations for $\alpha$
and $\beta$ introduced earlier in Figs.~\fone\ and \ftwo), where
the sense of the crossing of lines indicates which way the
corresponding blocks are analytically continued in the corresponding
variables.  Then Fig.~\ffive\ illustrates the relation
  $$
  \beta^\pm_{ij}(a,b,c,d)=\alpha_{ik}(a,b,c,d)\gamma^\pm_k(b,c)
  \alpha_{kj}(c,b,d,a).
  \eqn\Vbeaga
  $$
A similar and simpler argument gives
  $$
  \delta_{ij}=\alpha_{ik}(a,b,c,d)\alpha_{kj}(b,c,d,a).
  \eqn\Voeas
  $$
Since we are interested in the braiding properties of the
\FS\ current, we will take all external legs to be the
\JK\ field, of dimension $\Delta=(K+4)/(K+2)$.
The $\gamma^\pm_j$ are then given by
  $$
  \gamma^\pm=\left\{{\rm exp}\left(\pm{{2\pi i}\over{K+2}}\right),
  -{\rm exp}\left(\mp{{2\pi i}\over{K+2}}\right),
  {\rm exp}\left(\mp{{4\pi i}\over{K+2}}\right)\right\}
  \eqn\Vgd
  $$
where the last element, $\gamma_3^\pm$, corresponds to two \JK\
currents fusing to the identity, the middle to \JK, and the first
to the extra chiral field (called $\phi_{5,1}$ in
Section~1) that should decouple by the closure condition.
We can drop the external leg arguments of $\alpha$, $\beta^\pm$,
and $\gamma^\pm$, and write, as a simple consequence of
\Vbeaga\ and \Voeas
  $$
  (\beta^\pm)^n=\alpha(\gamma^\pm)^n\alpha,
  \eqn\Vbn
  $$
where $\gamma^\pm$ is to be thought of as a diagonal matrix, and
matrix multiplication is understood.  Since correlation
functions of \JK\ currents are invariant under $\alpha$
transformations (this is the associativity condition), they
will also be invariant under $(\beta^\pm)^n$ by \Vbn\ if
$n$ is such that $(\gamma^\pm)^n=1$.  From \Vgd\ we see
that if $K$ is even this is achieved for $n=K+2$, and if
$K$ is odd, for $n=2(K+2)$.

However, from the discussion
of Section~1 we recall that the \JK\ correlation
functions satisfy not only the associativity condition but
also the closure condition which states that no other
fields than $T$ and \JK\ enter into the \AK\ \ca.  In
particular, the $\phi_{5,1}$ field decouples, so the top
row and first column of the $\alpha$, $\beta^\pm$, and
$\gamma^\pm$ matrices play no role.  This means that
the \JK\ four-point function will be invariant under
$(\beta^\pm)^n$ transformations for $n$ such that
the lower two entries in $(\gamma^\pm)^n$ are one.
This permits the stronger statement that, for the
case where $K=4m$ for some integer $m$, the \JK\
correlator is invariant under $(\beta^\pm)^n$ with
$n=(K+2)/2$.  In this case $n$ is odd, so this is a
condition involving braidings in which the \JK\ fields
are interchanged an odd number of times, and so can
be thought of as a non-trivial generalization of the
notion of statistics for the \JK\ currents.  It is
important to note that these statistics are not just
the Abelian phases of ``fractional statistics.''
Indeed, under a single interchange of two \JK\ fields,
we obtain two new fields related to the old \JK\
fields not by an overall phase, but by the action of the
matrix $\beta^\pm$ on their chiral vertices.

In fact, using the explicit form for $\beta^\pm$ derived
in Appendix~A, one can show that demanding invariance
of the \JK\ correlator when $K=4m$ under
$(\beta^\pm)^{(K+2)/2}$ transformations, as well as
under fusion ($\alpha$) transformations, is equivalent
to demanding the closure condition.  This suggests that
the algebras with $K=4m$ can be characterized as those
that are invariant under the subgroup of the full
transformation group generated by $\alpha$ and
$(\beta^\pm)^{(K+2)/2}$.  We do not know whether a similar
characterization exists for the $K\neq 4m$ theories.

\noindent {\bf b.}~~~One reason for considering extended
Virasoro algebras like \AK\ is that infinite numbers
of Virasoro primaries are organized into extended \ca\
descendants of a smaller number of \ca\ primary fields.
For this reason it is important to look for extended
algebras which are in some sense maximal.  So far, we
have considered the case where the Virasoro algebra is
extended by only one additional current \JK.  For the
spin-$4/3$ ($K=4$) case, Fateev and Zamolodchikov [\ZFft]
have constructed an extended algebra with two additional
currents of the same dimension $\Delta=(K+4)/(K+2)$:
  $$\eqalign{
  \psi(z)\psi(w)= & \sqrt{2}\lambda_4(z-w)^{-\Delta}
  \left\{\psi^{\dagger}(w)+\ldots\right\},\cr
  \psi^{\dagger}(z)\psi^{\dagger}(w)= & \sqrt{2}\lambda_4
  (z-w)^{-\Delta}\left\{\psi(w)+\ldots\right\},\cr
  \psi(z)\psi^{\dagger}(w)= & (z-w)^{-2\Delta}\left\{1+
  {2\Delta\over c}(z-w)^2T(w)+\ldots\right\}.\cr}
  \eqn\Vpsi
  $$
Here $\lambda_4$ is given by \Zii\ and \Zia\ with $K=4$,
implying that $(c_{111})^2=1$ and
  $$
  \lambda_4^2={2\over9}\left[{8\over c}-1\right].
  \eqn\Vlamb
  $$
Note that $J^{(4)}=(\psi+\psi^\dagger)/\sqrt{2}$ satisfies the
${\cal A}^{(4)}$ algebra \Zi.  An important property of this
splitting of the ${\cal A}^{(4)}$ algebra is that there is only
a single cut appearing on the right hand side of the OPEs \Vpsi.
This implies that the $\psi(z)$ and $\psi^\dagger(z)$ fields
obey Abelian braid relations, and simplifies the analysis of the
spin-$4/3$ algebra considerably.

The existence of this splitting of the ${\cal A}^{(4)}$ algebra
might have been guessed on other grounds.
Let us consider the minimal model with $c=6/7$ ($K=4$), the
tricritical three-state Potts model [\FQS], which we have
seen is a representation of the ${\cal A}^{(4)}$ \ca.
Its Virasoro primary fields may be labelled by two
indices $\phi_{m,n}$.  When we enlarge the Virasoro
symmetry to the ${\cal A}^{(4)}=\{J,T\}$ or the $\{\psi,\psi^\dagger,T\}$
extended algebra, some of the $\phi_{m,n}$ become
descendants with respect to the extended algebra
of other Virasoro primary fields.  Specifically,
$\phi_{3,1}$ becomes a descendant of the identity
$\phi_{1,1}$.  In the \FS\ algebra \Zi, there is one
$\phi_{3,1}$ field:  $\phi_{3,1}=J_{-\Delta}\phi_{1,1}$.
In Fateev and Zamolodchikov's spin-$4/3$ algebra, there
are two $\phi_{3,1}$ fields: $\psi_{-\Delta}\phi_{1,1}$
and $\psi^\dagger_{-\Delta}\phi_{1,1}$.  Let us
denote the characters corresponding to the Virasoro
primaries $\phi_{m,n}$ by $\chi_{m,n}$.  Then the
spin-$4/3$ extended algebra characters are given by
  $$
  {\rm ch}_n=\chi_{1,n}+2\chi_{3,n}+\chi_{5,n}.
  \eqn\Vch
  $$
The resulting modular invariant partition function
${\cal Z}(D_4)$ corresponds to the $(D_4,A_6)$ case
(or the $D$ case) in the $ADE$ classification [\CIZ].
This argument easily generalizes to all the
$SU(2)_4\otimes SU(2)_L/SU(2)_{4+L}$ coset models
(where the Virasoro characters are generalized to the
appropriate branching functions [\RavMod]).

This picture and the $ADE$ classification of modular invariant partition
functions [\CIZ] suggests that when $K=4m$ there exist new symmetry
algebras differing from the fractional supersymmetries \AK\ by having
additional fractional spin currents entering into the \ca, similar to
\Vpsi.  To simplify the analysis, the goal is to enlarge these algebras
so that they will satisfy Abelian braiding relations.

\noindent {\bf c.}~~~The associativity analysis for extended
Virasoro algebras in the minimal models can be applied to
currents other than the $\phi_{3,1}$ field, which was the
one considered in Section~2.  For simplicity, let us consider
chiral algebras involving the energy-momentum tensor and
just one other current, $\phi_{M,1}$.  From the minimal
model fusion rules, the fusion matrix, $\alpha_M$, for the
four-point function of this current is an $M\times M$
matrix.  By \Voeas, $(\alpha_M)^2=1$, so all its eigenvalues
are $\pm1$.  Recall that the associativity condition
on our chiral algebra states that the vector of structure
constants of the algebra is an eigenvector with eigenvalue
$+1$ of $\alpha_M$.  Let $n$ denote the number of $+1$
eigenvalues.  Then there will be an $n-1$ dimensional space
of solutions to the associativity conditions. (One entry
of the vector of structure constants is fixed by the
normalization of the $\phi_{M,1}$ field.)

However, for our proposed chiral algebra to have only $T$
and $\phi_{M,1}$ as chiral currents, there must be more
constraints to prevent other fields from coupling.  These
are just the closure conditions for the chiral algebra.  In
particular, the $\phi_{M,1}\phi_{M,1}$ OPE closes on the set
$\{\phi_{1,1},\ \phi_{3,1},\ \ldots,\ \phi_{2M-1,1}\}$, so
there are $M-2$ (if $M$ is odd) or $M-1$ (if $M$ is even)
further conditions on the vector of structure constants needed
to decouple the unwanted fields.  This means that, generically,
for there to be a solution to the associativity and closure
conditions, we must have $n\geq M-1$ (for $M$ odd) or $n=M$
(for $M$ even).  If $n=M$, then $\alpha_M$ would be the identity
matrix, which is not the case.  In fact, using the result for the
fusion matrix given in Appendix~A (A11), we find that for $M=2m$
the number of $+1$ eigenvalues is $m$, while for $M=2m+1$
the number of $+1$ eigenvalues is $m+1$.  Thus the only generic
solution to the associativity and closure conditions occurs
for $M=3$, which correspond to the \FS\ \ca s \AK\ we have been
studying in this paper.

Though the above argument was presented for the minimal models,
it only depended on the eigenvalues of the fusion matrix.
It should be clear from the discussions of previous sections
that this argument is therefore valid for the whole range of
$SU(2)_K\otimes SU(2)_L/SU(2)_{K+L}$ coset representations.

This work was supported in part by the National Science Foundation.


\Appendix{A}

This Appendix presents explicit formulas for the fusion and braiding
matrices for conformal blocks of the $c<1$ minimal models.  These
matrices, or at least parts of them, have been calculated in
different ways in ref.~[\DotFat,\FFK].  We will follow Dotsenko
and Fateev's method in what follows; for more details, the reader
is referred to ref.~[\DotFat].

The unitary minimal models have central charge given in \IIi,
and primary fields\foot{We have used the opposite convention of [\DotFat]
so that our $\phi_{n,m}$ is their $\phi_{m,n}$.} $\phi_{n,m}$ where $m,n$
are integers satisfying $1 \leq n \leq K+1,$ $1 \leq m \leq n.$ We
specialize to the subalgebra with $m=1,$ since that is all we use in the
rest of this paper.  The conformal dimension of the $\phi_{n,1}$ field is
$\Delta_{n,1} = \coeff{1}{4}(n-1)[n-1+(n+1)/(K+2)]$.

Dotsenko and Fateev derive integral expressions for the conformal blocks
of four-point functions
using a Feigin-Fuchs [\FeiF] construction of the minimal models as
a boson with background charge. We omit the details of introducing
conjugate fields and screening charges in what follows.
They show that there are $n$ independent
conformal blocks, where $n=(n_1+n_2+n_3-n_4)/2$, so that
  $$\eqalign{
  G(z_i) & = \langle \phi_{n_1,1} (z_1) \phi_{n_2,1} (z_2)
  \phi_{n_3,1} (z_3) \phi_{n_4,1} (z_4)
  \rangle \cr
  & = \sum^n_{k=1} A^{(n)}_k (a,b,c;\tilde\rho)
  F^{(n)}_k(a,b,c;\tilde\rho;z_i),\cr}
  \eqn\IIxiv
  $$
where we have defined the useful parameters
  $$\matrix{
  a = 1 - n_1, & b = 1 - n_2, \cr
  c = 1 - n_3, & \tilde\rho = 1 + {1 \over K+2}. \cr}
  \eqn\IIxiii
  $$
Without loss of generality we have taken $n_4\geq n_1,n_2,n_3.$
The $A^{(n)}_k (a,b,c;\tilde\rho)$ in \IIxiv\ are the \sc s to
be determined by the associativity conditions explained in Section~2.
The expression for the correctly normalized conformal blocks is
  $$\eqalign{
  F^{(n)}_k(a,b,c;\tilde\rho;z_i)=
  & z_{13}^{-\Delta_1 -\Delta_2 -\Delta_3 +\Delta_4}
    z_{14}^{-\Delta_1 +\Delta_2 +\Delta_3 -\Delta_4}
    z_{34}^{\Delta_1 +\Delta_2 -\Delta_3 -\Delta_4}
    z_{24}^{-2\Delta_2} \cr
  & \times \eta^{ab\tilde\rho/2}{(1-\eta)}^{bc\tilde\rho/2}
    \left\{N^{(n)}_k (a, b, c; \tilde\rho)\right\}^{-1}
    I^{(n)}_k(a, b, c; \tilde\rho ; \eta), \cr}
  \eqn\Aixa
  $$
where we have defined the integrals
  $$\eqalign{
  I^{(n)}_k(a,b,c;\tilde\rho;\eta)=
  & \int_1^\infty\!\! du_1 \ldots\!\!\!\!
    \int_1^{u_{n-k-1}}\!\!\! du_{n-k}
    \int_0^\eta\!\! du_{n-k+1} \ldots\!\!\!
    \int_0^{u_{n-2}}\!\!\! du_{n-1}
    \prod_{i<j}^{n-1} {(u_i-u_j)}^{2\tilde\rho} \cr
  & \times \prod_{i=1}^{n-k} u_i^{a\tilde\rho}
    {(u_i-\eta)}^{b\tilde\rho} {(u_i-1)}^{c\tilde\rho}
    \prod_{i=n-k+1}^{n-1}\!\!\! u_i^{a\tilde\rho}
    {(\eta-u_i)}^{b\tilde\rho} {(1-u_i)}^{c\tilde\rho},\cr}
  \eqn\Aix
  $$
and the following normalization constants [\DotFat]
  $$\eqalign{
  N^{(n)}_k(a,b,c;\tilde\rho) =
  & \prod_{i=1}^{n-k}{{\Gamma(i\tilde\rho)}\over{\Gamma(\tilde\rho)}}
    \prod_{i=1}^{k-1}{{\Gamma(i\tilde\rho)}\over{\Gamma(\tilde\rho)}}
    \prod_{i=0}^{k-2}{{\Gamma(1+\tilde\rho(a+i))
    \Gamma(1+\tilde\rho(b+i))}\over{\Gamma(2+\tilde\rho(a+b+k-2+i))}} \cr
  & \times \prod_{i=0}^{n-k-1}{{\Gamma(-1-\tilde\rho(a+b+c+2n-4-i))
    \Gamma(1+\tilde\rho(c+i))}\over{\Gamma(-\tilde\rho(a+b+2k-2+i))}}. \cr}
  \eqn\Aixb
  $$
We use the notations $\Delta_i=\Delta_{n_i,1}$, $z_i-z_j=z_{ij}$,
$\eta=(z_{12}z_{34})/(z_{13}z_{24})$, and $\eta$ is taken to
lie on the real axis between $0$ and $1$. Other values of $\eta$
can be reached by analytic continuation.  These formulas describe
conformal blocks which are correctly normalized for the field
at $z_1$ fusing with the field at $z_2$.  This follows from the
fact that in the $z_1\rightarrow z_2$ ($\eta\rightarrow 0$) limit
  $$
  I^{(n)}_k(a,b,c;\tilde\rho;\eta)\rightarrow
  N^{(n)}_k(a,b,c;\tilde\rho)\eta^{(k-1)(1+\tilde\rho(a+b+k-2))}f(\eta),
  \eqn\Aixc
  $$
where $f$ is some analytic function of $\eta$ which satisfies
$f(0)=1$.

The fusion matrix $\alpha$ relates the above conformal blocks to those
which are appropriately normalized as $z_2\rightarrow z_3$
($\eta\rightarrow 1$).  To calculate the relation between these two
sets of blocks, change variables in $I^{(n)}_k$ to $\tilde\eta=1-\eta$
and $\tilde u_i=1-u_i$, to find
  $$\eqalign{
  I^{(n)}_k(a,b,c;\tilde\rho;\eta)=
  & \int_{-\infty}^0\! d\tilde u_1\ldots\!\!\!
    \int_{\tilde u_{n-k-1}}^0\!\!\! d\tilde u_{n-k}
    \int_{\tilde\eta}^1\! d\tilde u_{n-k+1} \ldots\!\!\!
    \int_{\tilde u_{n-2}}^1 \! d\tilde u_{n-1}
    \prod_{i<j}^{n-1} (\tilde u_j-\tilde u_i)^{2 \tilde\rho} \cr
  & \times\prod_{i=1}^{n-k} (-\tilde u_i)^{c\tilde\rho}
    (\tilde \eta-\tilde u_i)^{b\tilde\rho}(1-\tilde u_i)^{a\tilde\rho}
    \prod_{i=n-k+1}^{n-1}\!\!\! \tilde u_i^{c\tilde\rho}
    (\tilde u_i-\tilde\eta)^{b\tilde\rho}(1-\tilde u_i)^{a\tilde\rho}.\cr}
  \eqn\Aiv
  $$
Notice that the integrands of \Aiv\ and \Aix\ are the same
except for the switch of $a$ and $c$; the difference
between the two expressions is the limits of integration. In the
complex plane, however, the only singularities are at 0, $\eta,$ 1 and
$\infty$ so that using contour deformation the integration limits in
\Aiv\ can be pulled over to match those in \Aix.  In this way we can
calculate the elements of the transformation matrix $\tilde\alpha$,
defined by
  $$
  I^{(n)}_k(a,b,c;\tilde\rho;\eta) =
  \tilde{\alpha}^{(n)}_{kj}(a,b,c;\tilde\rho)
  I^{(n)}_j(c,b,a;\tilde\rho;\tilde\eta=1-\eta).
  \eqn\Ai
  $$
The $I^{(n)}_j$'s on the right hand side are now in the correct
form for conformal blocks in the $\tilde\eta\rightarrow 0$ basis.
{}From the complete expression \Aixa\ for the conformal blocks and the
normalization of the $I^{(n)}_k$'s \Aixc, we see that the fusion
matrix $\alpha$ is given by
  $$
  \alpha^{(n)}_{kj}(a,b,c;\tilde\rho)=
  \left\{N^{(n)}_k(a,b,c;\tilde\rho)\right\}^{-1}
  \tilde{\alpha}^{(n)}_{kj}(a,b,c;\tilde\rho)N^{(n)}_j(c,b,a;\tilde\rho).
  \eqn\Aiva
  $$
Note that if $a=c$, $\alpha$ differs from $\tilde\alpha$ by a
similarity transformation and so has the same eigenvalues.

In ref.~[\DotFat] Dotsenko and Fateev
calculate one column of $\tilde\alpha$
using the above described contour manipulation
techniques.  We find that it is not difficult to calculate the
entire matrix using the same approach.
If we define the auxiliary variable
$\tilde x=e^{i\pi\tilde\rho}$, the
function $[d]=\tilde{x}^d-\tilde{x}^{-d}$,
and for integral $d$ and $d'$ the symbol
  $$
  \brak{d}{d'} = \left\{ \matrix{
  \prod^{d'}_{i=d} [i] & d' \geq d \cr
  1 & d' < d  \cr} \right.
  \eqn\Aii
  $$
then the explicit form for the $\tilde\alpha$ matrix is
  $$\eqalign{
  \tilde{\alpha}^{(n)}_{kj}(a,b,c;\tilde\rho) & =
  \sum^{{\rm min}(k-1,j-1)}_{i={\rm max}(0,k+j-n-1)}
  \!\!\!\!\!\!\!\!\!{(-1)}^{k+j-i}
  [b+c+2j-3] [b+c+n-k-1+2i] \cr
  & \times
  \brak{a+k-1-i}{a+n-j-1}
  \brak{b+i}{b+j-2}
  \brak{c+n-k}{c+n-k-1+i} \cr
  & \times
  \brak{a+b+c+n-2+i}{a+b+c+n+k-4}
  \brak{1+i}{j-1}
  \brak{k-i}{n-j} \cr
  & \times \Biggl\{
  \brak{b+c+n-k-1+i}{b+c+n-2+i}
  \brak{b+c+j-2+i}{b+c+n-k+j-2+i} \cr
  & \quad \times
  \brak{1}{j-1-i}
  \brak{1}{n-k-j+1+i}
  \Biggr\}^{-1}.\cr}
  \eqn\Aiii
  $$

Now we would like to consider the braiding matrix $\beta$ for four-point
blocks.  This matrix relates blocks properly normalized as
$z_1\rightarrow z_2$ to those normalized as $z_1 \rightarrow z_3$.
Since the operators are radially ordered, for $z_1$ to approach $z_3$
it must be continued around the operator at $z_2$.  In terms of the
integral representations of the conformal blocks, this corresponds to
continuing the parameter $\eta\rightarrow 1/\eta$.  Different
transformations result depending on whether we continue
above or below the singularity at $\eta=1$ in the complex
plane.  This analytic continuation can be performed explicitly
using expressions \Aixa--\Aixb\ for the conformal blocks.

However, it turns out [\MooSigh,\FFKI] that the braiding matrices are
simply expressible in terms of the fusion matrix and ``twisting''
matrices $\gamma^\pm$, introduced above in Section 5.
The relation we are interested in between $\alpha$, $\beta^\pm$,
and $\gamma^\pm$ is most easily described diagrammatically.
We adopt the notation for $\alpha$, $\beta$, and $\gamma$
introduced earlier in Figs.~\fone, \ftwo, and \ffour,
where the sense of the crossing of lines indicates which way the
corresponding blocks are analytically continued in the corresponding
variables.  Then Fig.~\fsix\ illustrates the relation
  $$
  \beta^\pm_{jk}(a,b,c,d)=\gamma^\mp_j(a,b)
  \alpha_{jk}(b,a,c,d)\gamma^\pm_d(k,b).
  \eqn\Vbegag
  $$
Note that the equivalence of \Vbegag\ with the alternate
expression for $\beta^\pm$ in terms of $\alpha$ and $\gamma^\pm$
given by \Vbeaga\ is just an expression of the Yang-Baxter
relation that braiding matrices must satisfy.
Translating into the parameters used above to describe the minimal
model conformal blocks, we derive from \Vbegag\ the simple formula for
the braiding matrix in terms of the fusion matrix
  $$\eqalign{
  \beta^{(n)\pm}_{jk}(a,b,c;\tilde\rho) = &
  {\rm e}^{\mp i\pi(\Delta_1+\Delta_2+\Delta_3-\Delta_4+j+k-2)}
  \alpha^{(n)}_{jk}(b,a,c;\tilde\rho) \cr
  & \times{\tilde x}^{\pm (1-j)(a+b+j-2)}
  {\tilde x}^{\pm (1-k)(a+c+k-2)}.\cr}
  \eqn\Abeta
  $$


\Appendix{B}

This Appendix describes the calculation of the OPEs of the
$Z_K$-parafermion fields $\epsilon_1$ and $\eta$, introduced
in Section 3.  These fields are parafermion descendants of the
$\sigma_2$ spin field. Since the $Z_K$-parafermion theory is
simply the $SU(2)_K/U(1)$ coset theory, our strategy will be to
express these fields in terms of $SU(2)_K$ WZW fields and currents,
and use the $SU(2)_K$ Ward identities to calculate their OPEs.
Since the $\sigma_2$ spin field is related to the WZW spin-$1$
primary field [\ZFPara], we will concentrate on the WZW spin-$1$ with
WZW spin-$1$ OPE.

The singular terms in the $SU(2)_K$ current-current OPE are
  $$\eqalign{
  J^+(z)J^-(w)= & {K\over(z-w)^2}+{2J^0(w)\over(z-w)},\cr
  J^0(z)J^\pm(w)= & \pm{J^\pm(w)\over(z-w)},\cr
  J^0(z)J^0(w)= & {K/2\over(z-w)^2}.\cr}
  \eqn\Bi
  $$
The spin-$j$ primaries $\Phi^{(j)}_m$ have dimension
$\Delta_j=j(j+1)/(K+2)$.  We choose their relative normalizations
so that
  $$\eqalign{
  J^0_0\,\Phi^{(j)}_m= & \ m\Phi^{(j)}_m,\cr
  J^\pm_0\,\Phi^{(j)}_m= & \ \sqrt{(j\pm m+1)(j\mp m)}
  \Phi^{(j)}_{m\pm1},\cr}
  \eqn\Bii
  $$
where $J^{\pm,0}_0$ are the zero modes of the Kac-Moody
currents.  We fix the overall normalization of the
spin-$1$ fields by
  $$
  \Phi^{(1)}_+(z) \Phi^{(1)}_-(0)= 1\cdot z^{-4/(K+2)}+\ldots
  \eqn\Biii
  $$
If the $Z_K$-parafermion currents $\psi_1$ and $\psi^+_1$ are
defined by
  $$\eqalign{
  J^+= & \ \sqrt{K}\psi_1\exp\{i\varphi\sqrt{2/K}\}\cr
  J^0= & \ i\sqrt{K/2}\partial\varphi\cr
  J^-= & \ \sqrt{K}\psi^+_1\exp\{-i\varphi\sqrt{2/K}\}\cr}
  \eqn\Biv
  $$
where $\varphi$ is a free boson satisfying
$\langle\varphi(z)\varphi(w)\rangle= -{\rm ln}(z-w)$, then the
parafermion fields are related to $\Phi^{(1)}_m$ by [\ZFPara]
  $$\eqalign{
  \Phi^{(1)}_+= & \ \sigma_2\exp\{i\varphi\sqrt{2/K}\}\cr
  \Phi^{(1)}_0= & \ \sqrt{K/2}\epsilon_1\cr
  \Phi^{(1)}_-= & \ \sigma^+_2\exp\{-i\varphi\sqrt{2/K}\}.\cr}
  \eqn\Bv
  $$
The coefficients in \Bv\ are determined by \Bii--\Biv, the
definition of $\epsilon_1$ \IViii, and the choice of normalization
of the spin fields:
  $$
  \sigma_2(z)\sigma^+_2(w)= 1\cdot z^{-{2(K-2)\over K(K+2)}}+\ldots
  \eqn\Bvi
  $$
{}From \Biv\ and \Bv, and the definition of $\eta=\hat\epsilon_1-
\hat\epsilon^+_1$, it is also easy to see that
  $$
  \eta={1\over\sqrt{K}}\left(J^-_{-1}\Phi^{(1)}_+ +
  {4\sqrt2\over K}J^0_{-1}\Phi^{(1)}_0 -
  J^+_{-1}\Phi^{(1)}_-\right).
  \eqn\Bvii
  $$

Following ref.~[\ZFWZW], introduce $SU(2)$-invariant combinations
depending on a new auxiliary parameter $x$
  $$\eqalign{
  J(x,z) = & J^-(z) + 2xJ^0(z)- x^2 J^+(z),\cr
  \Phi^{(j)}(x,z) = & \sum_{m=-j}^j \left(\matrix{2j\cr
  j+m\cr}\right)^{1/2} x^{j+m}\Phi^{(j)}_m(z),\cr}
  \eqn\Bviii
  $$
from which it follows with our normalization \Bii\ that
  $$\eqalign{
  J^+_0(x)\Phi^{(j)}(x,z)= & \partial_x \Phi^{(j)}(x,z)\cr
  J^0_0(x)\Phi^{(j)}(x,z)= & -j \Phi^{(j)}(x,z)\cr
  J^-_0(x)\Phi^{(j)}(x,z)= & 0. \cr}
  \eqn\Bix
  $$
We have introduced the deformed current modes $J^{\alpha}_m(x)$
which obey the usual Kac-Moody current mode commutation relations.
They are defined by
  $$\eqalign{
  J^+_n(x)= & J^+_n \cr
  J^0_n(x)= & J^0_n-xJ^+_n \cr
  J^-_n(x)= & J^-_n+2xJ^0_n-x^2 J^+_n .\cr}
  \eqn\Bx
  $$
In terms of these fields and modes, the spin-$j_1$ with spin-$j_2$
OPE can be written in a unified way as
  $$\eqalign{
  \Phi^{(j_1)}(x,z)\Phi^{(j_2)}(y,0)= & \sum_{j_3} c_{j_1j_2j_3}
  z^{\Delta_{j_3}-\Delta_{j_1}-\Delta_{j_2}}\sum_{m=0}^\infty z^m
  \sum_{\{\alpha_i\}}J^{\alpha_1}_{-1}(y)\ldots J^{\alpha_m}_{-1}(y)\cr
  &\times\sum_{k,l}C^{\{\alpha_i\}}_{k,l}(x-y)^{k+j_1+j_2-j_3}
  \left(J^+_0(y)\right)^l \Phi^{(j_3)}(y,0).\cr}
  \eqn\Bxi
  $$
The sum over $j_3$ runs over the spins allowed by the fusion rules,
and the normalization is set by $C_{0,0}=1$.  The structure constants
$c_{j_1j_2j_3}$ can be related to three-point functions by [\ZFWZW]
  $$\eqalign{
  \langle\Phi^{(j_1)}_{m_1}(z_1)
  \Phi^{(j_2)}_{m_2}(z_2)\Phi^{(j_3)}_{m_3}(z_3)
  \rangle = & [j_1j_2j_3]\left( \matrix{ j_1 & j_2 & j_3 \cr
  m_1 & m_2 & m_3 \cr} \right) c_{j_1j_2j_3} \cr
  & \times z_{12}^{\Delta_{j_3}-\Delta_{j_1}-\Delta_{j_2}}
  z_{13}^{\Delta_{j_2}-\Delta_{j_1}-\Delta_{j_3}}
  z_{23}^{\Delta_{j_1}-\Delta_{j_2}-\Delta_{j_3}}, \cr}
  \eqn\IIIi
  $$
where
  $$
  [j_1j_2j_3] = \sqrt{{(j_1+j_2-j_3)!(j_1-j_2+j_3)!(-j_1+j_2+j_3)!
  (j_1+j_2+j_3+1)!}\over {(2j_1)!(2j_2)!(2j_3)!}}
  \eqn\IIIii
  $$
and the second term in \IIIi\ is the standard 3-j symbol {[\Gott]}.

The coefficients $C^{\{\alpha_i\}}_{k,l}$ in \Bxi\ can be determined
recursively using the Kac-Moody Ward identities.  In particular, by
acting on both sides of \Bxi\ with $J^a_n(y,0)$ for $n=0,1$, we obtain
the relations, where $[\Phi^1\Phi^2]$ denotes the right hand side of \Bxi,
  $$\eqalign{
  J^+_0(y,0)[\Phi^1\Phi^2]= & (\partial_x+\partial_y)[\Phi^1\Phi^2]\cr
  J^0_0(y,0)[\Phi^1\Phi^2]= &
  \left((x-y)\partial_x-j_1-j_2\right)[\Phi^1\Phi^2]\cr
  J^-_0(y,0)[\Phi^1\Phi^2]= &
  (x-y)\left(2j_1-(x-y)\partial_x\right)[\Phi^1\Phi^2]\cr
  J^+_1(y,0)[\Phi^1\Phi^2]= & z\partial_x[\Phi^1\Phi^2]\cr
  J^0_1(y,0)[\Phi^1\Phi^2]= &
  z\left((x-y)\partial_x-j_1\right)[\Phi^1\Phi^2]\cr
  J^-_1(y,0)[\Phi^1\Phi^2]= &
  z(x-y)\left(2j_1-(x-y)\partial_x\right)[\Phi^1\Phi^2].\cr}
  \eqn\Bxii
  $$
These are sufficient to solve for the $C^{\{\alpha_i\}}_{k,l}$ using the
current algebra \Bi.  The result we find for the leading terms in the
$j_1=j_2=1$ OPE is
  $$\eqalign{
  \Phi^{(1)}(x,z)\Phi^{(1)}(y,0)= & c_{110}z^{-2\Delta_1}(x-y)^2\left\{1
  +{2z\over K}\left[J^0_{-1}(y)+(x-y)^{-1}J^-_{-1}(y)\right]
  +{\cal O}(z^2)\right\}\cr
  & +c_{111}z^{-\Delta_1}(x-y)\left\{\left[{(x-y)\over2}J^+_0(y)+1\right]
  \Phi^{(1)}(y,0)+{\cal O}(z)\right\}\cr
  & +c_{112}z^{\Delta_1} \left\{{\cal O}(1)\right\}.\cr}
  \eqn\Bxiii
  $$
Expanding according to \Bviii, we see that the normalization adopted
for the WZW spin-$1$ fields \Biii\ corresponds to setting $c_{110}=1$.

All the information we need to compute the leading terms of the
$\epsilon_1$ and $\eta$ OPEs is now at hand.  By \Bv, \Bvii, and
\Bix\ we can express the $\epsilon_1$ field in terms of the WZW
spin-$1$ primary as
  $$
  \epsilon_1=\sqrt{2\over K}\Phi^{(1)}_0
  = {1\over\sqrt{K}}\left[\partial_y\Phi^{(1)}(y,0)\right]\Bigr|_{y=0}
  = {1\over\sqrt{K}}\left[J^+_0(y)\Phi^{(1)}(y,0)\right]\Bigr|_{y=0}.
  \eqn\Bxiv
  $$
We use this expression in \Bxiii\ by taking $x$- and $y$-derivatives
of both sides, then evaluating at $x=y=0$. The leading term proportional
to $c_{111}$ vanishes, and the term proportional to $c_{110}$ contributes
  $$
  \epsilon_1(z)\epsilon_1(0)=-c_{110}\left({2\over K}\right)
  z^{-2\Delta+2}+c_{111}{\cal O}(z^{-\Delta+2})+\ldots
  \eqn\Bxv
  $$
where $\Delta=\Delta_1+1=(K+4)/(K+2)$.  The $\epsilon_1\eta$ OPE is
somewhat harder to evaluate.  From \Bvii\ we see that
  $$
  \eta={1\over{2\sqrt{K}}}\left(\partial_y^2+{4\over K}\partial_w
  \partial_y +\partial_w^2\right)\left[J_{-1}(w,0)\Phi^{(1)}(y,0)
  \right]\Bigr|_{x=y=0}.
  \eqn\Bxvi
  $$
A simple contour deformation argument using \Bxiv\ and \Bxvi\ shows
that
  $$\eqalign{
  \epsilon_1(z)\eta(0)= & {1\over 2K}\partial_x\left(\partial_y^2
  +{4\over K}\partial_w\partial_y+\partial_w^2\right)\cr
  &\times\left[J_{-1}(w)+{1\over z}\left(2(w-x) +(w-x)^2
  \partial_x\right)\right]\left(\Phi^{(1)}(x,z)\Phi^{(1)}(y,0)
  \right)\Bigr|_{x=y=w=0} \cr
  = & {1\over K}\left[{1\over z}(\partial^2_x-\partial^2_y)
  +{\cal O}(1)\right]\left(\Phi^{(1)}(x,z)\Phi^{(1)}(y,0)
  \right)\Bigr|_{x=y=0}.\cr}
  \eqn\Bxvii
  $$
In the second step we evaluated the $w$-derivatives, set
$w=0$, and kept only the leading power of $z$.  The leading term
proportional to $c_{110}$ in \Bxiii\ vanishes, and the term
proportional to $c_{111}$ contributes
  $$
  \epsilon_1(z)\eta(0)=c_{110}{\cal O}(z^{-2\Delta+2})+
  {{2c_{111}}\over\sqrt{K}}z^{-\Delta}\epsilon_1(0)+\ldots
  \eqn\Bxviii
  $$
Finally, the $\eta(z)\eta(0)$ OPE is evaluated in a similar
manner, using \Bxvi\ and a somewhat more complicated contour
manipulation, to obtain the leading $c_{110}$ term
  $$
  \eta(z)\eta(0)=-c_{110}\left({{2(K+4)(K-2)}\over{K^2}}\right)
  z^{-2\Delta}+\ldots
  \eqn\Bxix
  $$
The formulas \Bxv, \Bxviii, and \Bxix\ include only the minimum
number of terms in the $Z_K$-parafermion OPEs needed in Section 3
to evaluate the \lK\ coupling constant in the \JK\JK\
OPE.  There are other contributions to these OPEs which we have not
calculated in this Appendix.  Thus, for example, the
dots in \Bxix\ denote not only parafermion descendants of the
identity, but also parafermion descendants of the $\sigma_2$ spin
field (proportional to $c_{111}$) and the $\sigma_4$ spin field
(proportional to $c_{112}$).  It is a straightforward, if
laborious, exercise to calculate these terms along the lines
indicated above;  in fact, we have included the leading
$c_{111}$ terms of the $\epsilon_1\epsilon_1$ and $\eta\eta$
OPEs in (3.10).


\Appendix{C}

In this Appendix we assemble the integral expressions for the
conformal blocks of correlators of spin-$j$ primary fields of the
$SU(2)_K$ WZW model, as calculated in ref.~[\Dots].
The Wakimoto representation used to calculate these conformal
blocks is outlined in Section~4.

The spin-$j$ primary fields, $\Phi^{(j)}_m$, of the $SU(2)_K$
WZW model have dimensions $\Delta_j=j(j+1)/(K+2)$.
Since the $m$-dependence of the $\Phi^{(j)}_m$ OPEs is known from
$SU(2)_K$ Ward identities \IIIi, it is sufficient to consider
correlation functions of fields with $m=\pm j$.  In [\Dots]
correlators of two spin-$j$ with two spin-$l$ fields are
considered.  For $j>l$, it is shown that there are $n$
independent conformal blocks, where $n=2l+1$.  Thus
  $$\eqalign{
  {G}_{jl}(z_i) & =\langle\Phi^{(j)}_{-j}(z_1)\Phi^{(l)}_l(z_2)
  \Phi^{(l)}_{-l}(z_3)\Phi^{(j)}_j(z_4)\rangle\cr
  & = \sum_{k=1}^n A_k^{(n)}(a,b,c;\rho)
  F_k^{(n)}(a,b,c;\rho;z_i),\cr}
  \eqn\IIIx
  $$
where we have defined the parameters
  $$
  a=-2j,\quad\quad  b=c=-2l,\quad\quad {\rho}={1\over{K+2}}.
  \eqn\IIIxiii
  $$
The expression for the correctly normalized conformal blocks is
  $$\eqalign{
  F^{(n)}_k(a,b,c;{\rho};\eta)= &
   (z_{13}z_{24})^{-2\Delta_l}\,z_{14}^{2\Delta_l-2\Delta_j}\,
   \eta^{ab\rho/2}(1-\eta)^{bc\rho/2}\cr
  &\times\left\{N^{(n)}_k(R:a,b,c;{\rho})\right\}^{-1}
   I^{(n)}_k(R:a,b,c;{\rho};\eta),\cr}
  \eqn\IIIb
  $$
where we have defined the integrals
  $$\eqalign{
  I^{(n)}_k(R:a,b,c;\rho;\eta)= &
  \int_1^\infty\!du_1\ldots\!\!\!\!\!
  \int_1^{u_{n-k-1}}\!\!\!\!du_{n-k}\int_0^\eta\!du_{n-k+1}\ldots
  \!\!\!\!\int_0^{u_{n-2}}\!du_{n-1} R^{(n)}(u_i)
  \prod_{i<j}^{n-1}(u_i-u_j)^{2\rho}\cr
  & \times\prod_{i=1}^{n-k}u_i^{a\rho}(u_i-\eta)^{b\rho}
  (u_i-1)^{c\rho}\prod_{i=n-k+1}^{n-1}u_i^{a\rho}
  (\eta-u_i)^{b\rho}(1-u_i)^{c\rho},\cr}
  \eqn\IIIxiv
  $$
and the normalization constants [\Dots]
  $$\eqalign{
  N^{(n)}_k(R:a,b,c;{\rho}) = &
  (-1)^{n-k}{{(2j)!(n-k)!}\over{(2j-k+1)!}}
  \prod_{i=0}^{n-k-1}{{\Gamma(-{\rho}(a-2-i))
  \Gamma(1+{\rho}(c+i))}\over{\Gamma(1+{\rho}(-a+c+n-k+1+i))}}
  \cr & \times
  \prod_{i=0}^{k-2} {{\Gamma({\rho}(a+i))\Gamma(1+{\rho}(b+i))}
  \over{\Gamma(1+{\rho}(a+b+k-2+i))}}
  \prod_{i=1}^{n-k}{{\Gamma(i\rho)}\over{\Gamma(\rho)}}
  \prod_{i=1}^{k-1}{{\Gamma(i\rho)}\over{\Gamma(\rho)}}.\cr}
  \eqn\IIIxix
  $$
We use the notations, $z_i-z_j=z_{ij}$,
$\eta=(z_{12}z_{34})/(z_{13}z_{24})$, and $\eta$ is taken to
lie on the real axis between $0$ and $1$. Other values of $\eta$
can be reached by analytic continuation.  In \IIIxiv\
$R^{(n)}(u_i)$ is a rational function of the $n-1$ variables
$u_i$ given by
  $$
  R^{(n)}(u_i)=(n-1)!\sum_{p=0}^{n-1}\sum_{\{\sigma_i\}^p}
  \left(\matrix{2j\cr p\cr}\right)\left\{(u_1-\sigma_1)\ldots
  (u_{n-1}-\sigma_{n-1})\right\}^{-1},
  \eqn\IIIxv
  $$
where the second sum runs over all sets of parameters $\sigma_i$ such that
$p$ of them have the value 0 and $n-1-p$ of them have the value 1.
Note that when $R=1$, the above formulas for the integrals and their
normalizations reduce to the expressions (A4) and (A5) derived for the
minimal model in Appendix~A.

These formulas describe
conformal blocks which are correctly normalized for the field
at $z_1$ fusing with the field at $z_2$.  This follows from the
fact that in the $z_1\rightarrow z_2$ ($\eta\rightarrow 0$) limit
  $$
  I^{(n)}_k(R:a,b,c;\rho;\eta)\rightarrow
  N^{(n)}_k(R:a,b,c;\rho)
  \eta^{\Delta_{j+l-k+1}-\Delta_j-\Delta_l}f(\eta),
  \eqn\IIIxviii
  $$
where $f$ is some analytic function of $\eta$ which satisfies
$f(0)=1$.  The fusion matrix $\alpha$ relates the above conformal
blocks to those which are appropriately normalized as
$z_2\rightarrow z_3$ ($\eta\rightarrow1$).  We calculate the
relation between these two sets of blocks precisely as in the
minimal model case, explained in Appendix~A, by changing
variables in $I^{(n)}_k$ to $\tilde\eta=1-\eta$ and
$\tilde u_i=1-u_i$.  Since $R^{(n)}$ is invariant under these
changes of variables, we can write
  $$
  I^{(n)}_k(R:a,b,c;\rho;\eta)=
  \tilde{\alpha}^{(n)}_{kj}(a,b,c;\rho)
  I^{(n)}_j(R:c,b,a;\rho;1-\eta),
  \eqn\IIIxvi
  $$
where $\tilde\alpha^{(n)}_{kj}$ has the same form as found in
Appendix~A (A11) for the minimal series, but with
  $$
  {x}={\rm e}^{{\rm i}\pi{\rho}}=-\tilde x.
  \eqn\IIIxvii
  $$
in place of $\tilde x$.

{}From the complete expression \IIIb\ for the conformal blocks and the
normalization of the $I^{(n)}_k$'s \IIIxviii, we see that the fusion
matrix $\alpha$ is given by
  $$
  \alpha^{(n)}_{kj}(R:a,b,c;\rho)=
  \left\{N^{(n)}_k(R:a,b,c;\rho)\right\}^{-1}
  \tilde{\alpha}^{(n)}_{kj}(a,b,c;\rho)
  N^{(n)}_j(R:c,b,a;\rho).
  \eqn\Aiva
  $$
Note that if $a=c$, $\alpha$ differs from $\tilde\alpha$ by a
similarity transformation and so has the same eigenvalues.
When $a=c$, by (A11) $\tilde\alpha$ is purely even or odd
in $x$.  Thus the difference \IIIxvii\ between the $SU(2)_K$
and minimal model cases is at most an overall sign.  Recalling
the definitions of the parameters $a$, $b$, $c$, and $n$ for
the minimal model given in Appendix~A, we see that the fusion
matrix of the four-point function of spin-$j$ WZW primaries
differs from that of the $\phi_{2j+1,1}$ minimal model
primary field by a similarity transformation involving
the normalization constants.


\refout
\endpage
\tabout
\figout


\vbox{\offinterlineskip
\hrule
\halign{\strut#&\vrule#&~\hfil#\hfil~&
  \vrule#&~\hfil#\hfil~&\vrule#&~\hfil#\hfil~&
  \vrule#&~\hfil#\hfil~&\vrule#\cr
\omit&height8pt&\omit&&\omit&&\omit&&\omit&\cr
&&Representation of \AK
&&Central charge
&&`Free' field realization
&&\JK\ current&\cr
\omit&height6pt&\omit&&\omit&&\omit&&\omit&\cr
\noalign{\hrule}
\omit&height8pt&\omit&&\omit&&\omit&&\omit&\cr
&&$K{\rm th}$ minimal model
&&$c_{\rm min}={{K(K+5)}\over{(K+2)(K+3)}}$
&&$(\varphi,\alpha_0)$
&&$\phi_{3,1}$&\cr
\omit&height8pt&\omit&&\omit&&\omit&&\omit&\cr
&&${{SU(2)_K\otimes SU(2)_L}\over{SU(2)_{K+L}}}$ cosets
&&$c_{\rm min}\leq c\leq c_{SU(2)}$
&&$(\varphi,\alpha_0)+Z_K$-parafermion
&&$\epsilon_1\partial\varphi+\ldots$(3.7)&\cr
\omit&height10pt&\omit&&\omit&&\omit&&\omit&\cr
&&$SU(2)_K$ WZW model
&&$c_{SU(2)}={3K\over{K+2}}$
&&$(\varphi,\alpha_0)+(\omega,\omega^+)$
&&$q_{ab}J^a_{-1}\Phi^b_{(1)}$&\cr
\omit&height8pt&\omit&&\omit&&\omit&&\omit&\cr}
\hrule}
\bigskip
Table 1.

\end